\documentclass[twocolumn]{aastex631}
\let\tablenum\relax
\usepackage{siunitx}
\usepackage[version=4]{mhchem}
\usepackage{makecell}
\usepackage{amsmath}

\usepackage{listings}
\usepackage{color}
\usepackage{xcolor}
\usepackage{xparse}
\usepackage{graphicx}
\makeatletter
\@namedef{ver@ltcaption.sty}{2099/01/01 v9.99 dummy}
\providecommand*\LT@listentry[2]{}
\makeatother
\usepackage{caption}
\usepackage{subcaption}
\usepackage{xurl}

\newcommand\HII{$\textrm{H}\scriptstyle\mathrm{II}$}

\NewDocumentCommand{\codeword}{v}{
\texttt{\textcolor{blue}{#1}}
}
\definecolor{codegreen}{rgb}{0,0.6,0}
\definecolor{codegray}{rgb}{0.5,0.5,0.5}
\definecolor{codepurple}{rgb}{0.58,0,0.82}
\lstdefinestyle{mystyle}{
    commentstyle=\color{codegreen},
    keywordstyle=\color{magenta},
    numberstyle=\tiny\color{codegray},
    stringstyle=\color{codepurple},
    basicstyle=\ttfamily\footnotesize,
    breakatwhitespace=false,         
    breaklines=true,                 
    captionpos=b,                    
    keepspaces=true,                 
    numbers=left,                    
    numbersep=5pt,                  
    showspaces=false,                
    showstringspaces=false,
    showtabs=false,                  
    tabsize=2
}
\makeatletter
\AtBeginDocument{
  \def\footnoterule{
    \kern -3pt
    \hrule width 0.4\columnwidth
    \kern 2.6pt}
  \def\@makefntext#1{
    \hsize=\columnwidth
    \noindent
    \makebox[1.5em][l]{\textsuperscript{\@thefnmark}}
    \footnotesize #1}
}
\makeatother

\begin{document}
\title{VLA Observations of the Ionized Jet IRAS 16562$-$3959}
\author[0000-0002-9752-2125]{Ananay Sethi}
\affiliation{New Mexico Institute of Mining and Technology, Socorro, NM, USA}
\author[0000-0001-9942-7343]{Peter Hofner}
\affiliation{New Mexico Institute of Mining and Technology, Socorro, NM, USA}
\affiliation{Adjunct Astronomer at the National Radio Astronomy Observatory, Socorro, NM, USA}
\author[0000-0003-3168-5922]{Emmanuel Momjian}
\affiliation{National Radio Astronomy Observatory, Socorro, NM, USA}
\author[0000-0001-6755-9106]{Esteban D. Araya}
\affiliation{New Mexico Institute of Mining and Technology, Socorro, NM, USA}
\affiliation{George Mason University, Fairfax, VA, USA}
\author[0000-0003-0090-9137]{Tatiana M. Rodríguez}
\affiliation{University of Cologne, Cologne, Germany}
\author[0000-0003-4444-5602]{Stanley E. Kurtz}
\affiliation{Institute of Radio Astronomy and Astrophysics, Universidad Nacional Autonoma de Mexico, Mexico}

\begin{abstract}
\small
We present new high-resolution, high-sensitivity Karl G. Jansky Very Large Array (VLA) A-configuration observations of the high-mass protostellar jet system IRAS~16562$-$3959 at 3.1 and 6.1~GHz.
Our images resolve the five previously known jet components (OE, IE, C, IW, OW) into 21 distinct knots and sub-components, and reveal a new off-axis source, MW. 
Based on the dynamical timescales from previous works ($\tau_{\rm jet} \approx 620$ yr), the spacing and structure of the knots suggests episodic accretion with a mean interval between ejections $\sim 32$ yr.
We find that source MW has a substantial increase in peak brightness relative to its marginal detections in previous observations. 
MW exhibits a spectral index of $\alpha \approx -0.7$, characteristic of synchrotron emission from shock-accelerated relativistic electrons. 
Equipartition analysis yields a minimum-energy magnetic field strength $B_{\rm me} \approx 0.4$~mG, and a minimum total energy $E_{\rm me} \approx 4.4 \times 10^{42}$~erg. Archival 2MASS K$_s$ data imply a total outflow dynamical age of $\sim1{,}400$~yr, and together with a HST [FeII] $1.64\,\mu$m map, the NIR data
indicate that the fast jet is embedded in a wide angle wind with a $\sim 36^\circ$ opening angle.

\end{abstract}

\section{Introduction}\label{intro} 

High-mass stars ($M > 8\,M_\odot$) play an important role in shaping the interstellar medium (ISM) and driving galactic evolution. 
Through ionizing radiation and powerful stellar winds, they are the dominant sources of radiative and mechanical feedback to the ISM on galactic scales \citep{Zinnecker2007, Tan2014}, and through supernovae they are the primary factories of heavy elements in the Universe. 
Despite their importance, the physical processes governing high-mass star formation remain yet to be fully understood relative to their low-mass counterparts due to the following observational challenges:
First, high-mass star-forming regions are embedded in dense molecular clumps with typical dust extinction of $A_V \gg 100$, rendering the central protostellar objects inaccessible at optical and near-infrared wavelengths during the earliest, most active accretion phases. 
Second, high-mass protostars are intrinsically rare, and with them being located typically at distances of several kiloparsecs, sub-arcsecond-scale angular resolution is required to resolve the inner jet and disk-launching structures.  
These limitations have historically made it difficult to measure the key physical parameters such as accretion rate, outflow momentum, and jet launching radius that serve as observational constraints of theoretical models.

A central theoretical challenge in understanding high-mass star formation is the radiation pressure problem: as a massive protostar grows, its luminosity increases steeply ($L \propto M^3$), and early analytic estimates suggested that radiation pressure on infalling dust grains would halt accretion below the observed upper stellar mass limit \citep{Wolfire1987}. 
Modern simulations resolve the radiation pressure problem by demonstrating that disk-mediated accretion, in which material funnels through a circumstellar disk, allows radiation to escape along the polar directions \citep{Yorke2002, Krumholz2009}. 
This model explains ionized jets by launching matter magneto-centrifugally from the inner disk or disk--star interface \citep{Blandford1982, Ferreira2006}, transporting angular momentum outward, and thereby allowing accretion to proceed.
The detection of a collimated ionized jet associated with a high-mass protostar therefore constitutes indirect evidence for disk accretion, linking jet observations to the fundamental theoretical problem. 

The jet's spectral index $\alpha$ (where $S_\nu \propto \nu^\alpha$) serves as a key diagnostic of the emission mechanism.
Steep negative spectral indices, $\alpha \lesssim -0.5$, are a hallmark of non-thermal radiation produced when strong shocks accelerate particles via the diffusive shock acceleration mechanism \citep{Blandford1982}. 
By contrast, $\alpha \geq 0.6$ is expected for thermal free-free emission from a collimated, fully ionized jet. 
The coexistence of thermal and non-thermal emission within a single protostellar jet has been confirmed in only a small number of systems, with the clearest cases being HH~80--81 \citep{CarrascoGonzalez2010}, Cepheus~A HW-2 \citep{Rodriguez1994}, and the Serpens Triple Radio Source \citep{RodriguezKamenetzky2016}. 
The rarity of confirmed detections makes each new case scientifically significant as a constraint on shock acceleration efficiency and magnetic field amplification in the jet environment \citep{Araudo2021}. 

The large-scale geometry of a bipolar jet encodes additional physical information, about both the driving source and the medium through which the jet propagates. 
In some systems the two lobes deviate from strict bipolarity, which often is attributed either to precession of the jet axis, driven by a companion or a warped inner disk, or to the presence of a wide-angle outflow component surrounding the collimated jet. 
\citet{RodriguezKamenetzky2022} demonstrates that both a collimated jet and a wide-angle wind can be simultaneously active in high-mass protostellar systems, likely launched by X-wind or disk-wind mechanisms. Asymmetries can also be imposed by the external medium, even when the ejection itself is intrinsically symmetric. \citet{Gouveia1993} demonstrate that large scale density gradients, clumps, and cavities in the cloud, when interacting with the jet, can produce lobes of unequal length, brightness, and shock excitation. Collisions with dense clumps, or ram pressure from bulk motions of the ambient gas, can also deflect a jet, producing departures from bipolar symmetry without any change in the orientation of the launching axis.
Distinguishing between these scenarios requires a combination of radio morphological analysis, multi-epoch kinematic data, complementary infrared tracing of the outflow cavity walls, and independent maps of the ambient gas, such as molecular-line or dust continuum emission, to test for density asymmetries.

IRAS~16562$-$3959 (hereafter I16562) is a high-mass star-forming region at a distance of 1.8\,kpc, with an estimated mass $M \sim 10^3\,M_\odot$ derived from single dish 1.2$\,$mm dust observations, and a luminosity of $5.3 \times 10^4\,L_\odot$, obtained from SED modeling
of IRAS, MSX and $1.2\,$mm data \citep{Lopez2011}.
The IRAS source harbors a central High Mass Young Stellar Object (HMYSO) of likely mass $\sim 15\,$M$_\odot$ identified as G345.4938+01.4677 \citep{Guzman2014, e_guzman_2016}.

Subsequent Atacama Large Millimeter Array (ALMA) observations by \citet{Guzman2014} detected the central component of I16562 in several hydrogen recombination lines
which led these authors to propose a model of a slow photo-ionized wind with no need for a high-velocity component. 
However, a second epoch ATCA observation by \citet{e_guzman_2016} detected bipolar proper motion away from the central source, implying plane-of-sky velocities of $\sim 400 - 500\, \text{km s}^{-1}$ for the outer components, suggesting that
a fast jet emanates from the hypercompact\,\HII \ (HC\,\HII) region associated with the central source.
This detection of proper motion for several jet components constrained the dynamical age of the jet to about $620\,$yr.

\citet{Guzman2011} reported CO 3-2, 6-5, and 7-6 observations with APEX toward I16562 which showed the presence of a bipolar
molecular flow centered on the ionized jet, and oriented along the jet direction. The measured mass and momentum rate of the 
molecular flow are $1.92\,$M$_\odot$ and $3.0\times 10^{-2}\,$M$_\odot\,$km$\,$s$^{-1}\,$yr$^{-1}$, respectively. These values are typical
for flows from high-mass protostars, and there can be little doubt that the ionized jet is driving the extended molecular flow.

\citet{montes_chandra_2020} characterized the stellar population in the I16562 field using Chandra X-ray observations and near- to mid-infrared data from the VISTA VVV and Spitzer GLIMPSE surveys. Their observations showed that the high-mass protostar is located at the center of a young and dense cluster of lower mass stars, and is associated with an X-ray source.

Together with the aforementioned systems of Cep~A HW-2, HH~80-81, Serpens TRS, as well as IRAS~$16547-4247$ \citep{Rodriguez2005}, I16562 is part of a small group of rare high-mass protostellar jets characterized at radio wavelengths in sufficient detail to constrain jet physics, making it a key case study for understanding the relationship between accretion, outflow, and feedback in high-mass star formation. 
However, the previous ATCA observations, with typical synthesized beams of $\sim 2$--$3''$, lacked the angular resolution to resolve the inner jet into individual knots or to reliably decompose spatially blended components for spectral analysis.

In this paper we present new observations of I16562, providing a factor of $\sim 2.5$ improvement in angular resolution and $\sim 3\times$ improvement in sensitivity over previous radio images. 
We investigate the detailed morphology of the ionized jet, the presence of thermal and non-thermal emission, as well as flux variability and proper motion over the available three epochs, and then discuss the implications of our findings at the end of this paper. 

\section{Observations and Data Reduction}\label{observations}

We observed I16562 in the radio continuum on January 21 and 22, 2017 (project code VLA 16B-133) using the Karl G. Jansky Very Large Array (VLA)\footnote{The National Radio Astronomy Observatory and Green Bank Observatory are facilities of the U.S. National Science Foundation operated under cooperative agreement by Associated Universities, Inc.} in its A-configuration (maximum baseline 36 km). 
Observations were conducted in two frequency ranges: 2–4 GHz (VLA S-band) and 4–8 GHz (VLA C-band), with 8-bit and 3-bit samplers, respectively. 
The correlator setup fully covered each band, with 16 spectral windows (SPWs) in S-band and 32 SPWs in C-band, each of which had a bandwidth of 128 MHz. 
Throughout this paper we refer to the S- and C-band data as the 3.1 and 6.1 GHz observations, respectively, as these are the effective central frequencies of the respective bands of usable data. 
The 3.1 GHz setup used 64 channels per SPW with a channel width of 2 MHz, while the 6.1 GHz setup used 128 channels per SPW with a channel width of 1 MHz. 
The phase, delay, and pointing center were set at RA = $16^{\mathrm{h}}59^{\mathrm{m}}41^{\mathrm{s}}.63$, Dec = $-40^{\circ}03^{\prime}43^{\prime\prime}.6$ (J2000), corresponding to the 8.6 GHz peak position of the central source reported by \citet{e_guzman_string_2010}. 
The total on-source integration time was $\sim 30$ minutes in both bands, divided into 9-minute scans at 3.1 GHz and 5.5-minute scans at 6.1 GHz.
J1744–3116 served as the complex gain calibrator at 6.1 GHz, while J1626–2951 was used at 3.1 GHz. 
Flux density scale and bandpass calibration at both frequencies were done with J1331+3030 (3C286).

Data calibration was performed using the Common Astronomy Software Applications\footnote{\url{https://casa.nrao.edu/index.shtml}} (CASA) pipeline version 5.3.0-143, followed by imaging with the parallel-processing implementation of CASA (version 6.6.4) provided by the NRAO. 
Owing to the low elevation of the target, obstruction of certain baselines by local topography, intermittent antenna failures, and cloudy conditions, the standard continuum pipeline required significant manual intervention for flagging and calibration. 
Several SPWs were flagged due to these issues, and the resulting effective bandwidths, central frequencies, and SPW counts are summarized in Table \ref{tab:obsprop}.  
In addition to the initial calibration, we performed both phase-only and amplitude–phase self-calibration on the 3.1 and 6.1 GHz data sets independently to correct for residual antenna-based errors. 
This substantially improved image fidelity, and recovered weak emission structures. 

Imaging was carried out using the \codeword{tclean} task with a robust weighting parameter of 0.5, yielding the synthesized beam parameters listed in Table \ref{tab:obsprop}. 
The final images cover a $2^{\prime}\times2^{\prime}$ field with pixel scales of $0.130^{\prime\prime}$ and $0.066^{\prime\prime}$ for 3.1 and 6.1 GHz, respectively.  
Deconvolution was performed with the Multi-Term Multi-Frequency Synthesis (MTMFS) algorithm \citep{rau_cornwell_mtmfs_2011}, which models complex emission structures with a frequency-dependent Taylor expansion. 
We set \codeword{nterms} = 2 to simultaneously fit for continuum intensity and spectral index. 
For the 6.1 GHz data, we additionally adopted a multi-scale cleaning approach with \codeword{scales} = [0, 20, 40] pixels, enabling recovery of both compact and extended emission. 
The final rms noise values achieved after cleaning, along with the full observational parameters, are listed in Table \ref{tab:obsprop}.

\begin{deluxetable*}{cccccc}[h]
    \tabletypesize{\small}
    \tablewidth{0pt} 
    \tablecaption{Properties of Observations for I16562
    \label{tab:obsprop}}
    \tablehead{
        \colhead{\makecell{Frequency \\ band}} & 
        \colhead{\makecell{Effective \\ bandwidth}} & 
        \colhead{\makecell{Effective \\ Central \\ Frequency}} &
        \colhead{\makecell{Effective \\ SPWs}} & 
        \colhead{\makecell{Image Noise}} & 
        \colhead{$\theta_{\text{FWHM}}$} \\
        \colhead{} & 
        \colhead{(GHz)} & 
        \colhead{(GHz)} & 
        \colhead{} & 
        \colhead{($\mu$Jy beam$^{-1}$)} & 
        \colhead{($\theta_{\max}'' , \theta_{\min}''$, PA$^\circ$)}
    }
    \startdata
        S & 1.75 & 3.1 & 14 & 31 & 2.10, 0.48, $-$4.67 \\
        C & 3.80 & 6.1 & 27 & 6 & 1.03, 0.26, $-$3.43
    \enddata
\end{deluxetable*}

\begin{figure*}[!tbp]
     \centering
     \begin{subfigure}[!tbp]{\textwidth}
         \centering
         \includegraphics[width=\textwidth]{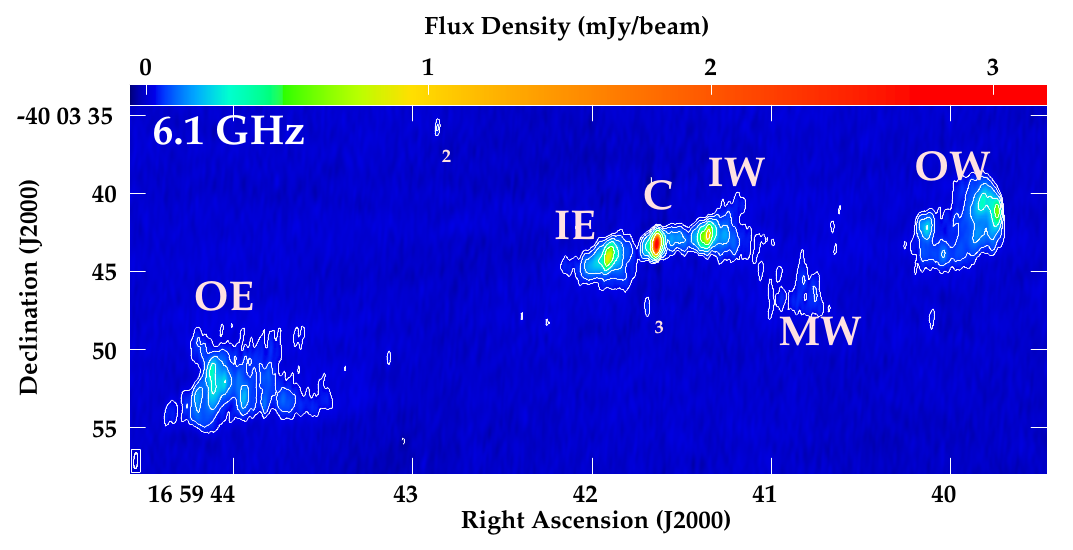}
         \label{fig:cband}
     \end{subfigure}
     \vfill
     \begin{subfigure}[!tbp]{\textwidth}
         \centering
         \includegraphics[width=\textwidth]{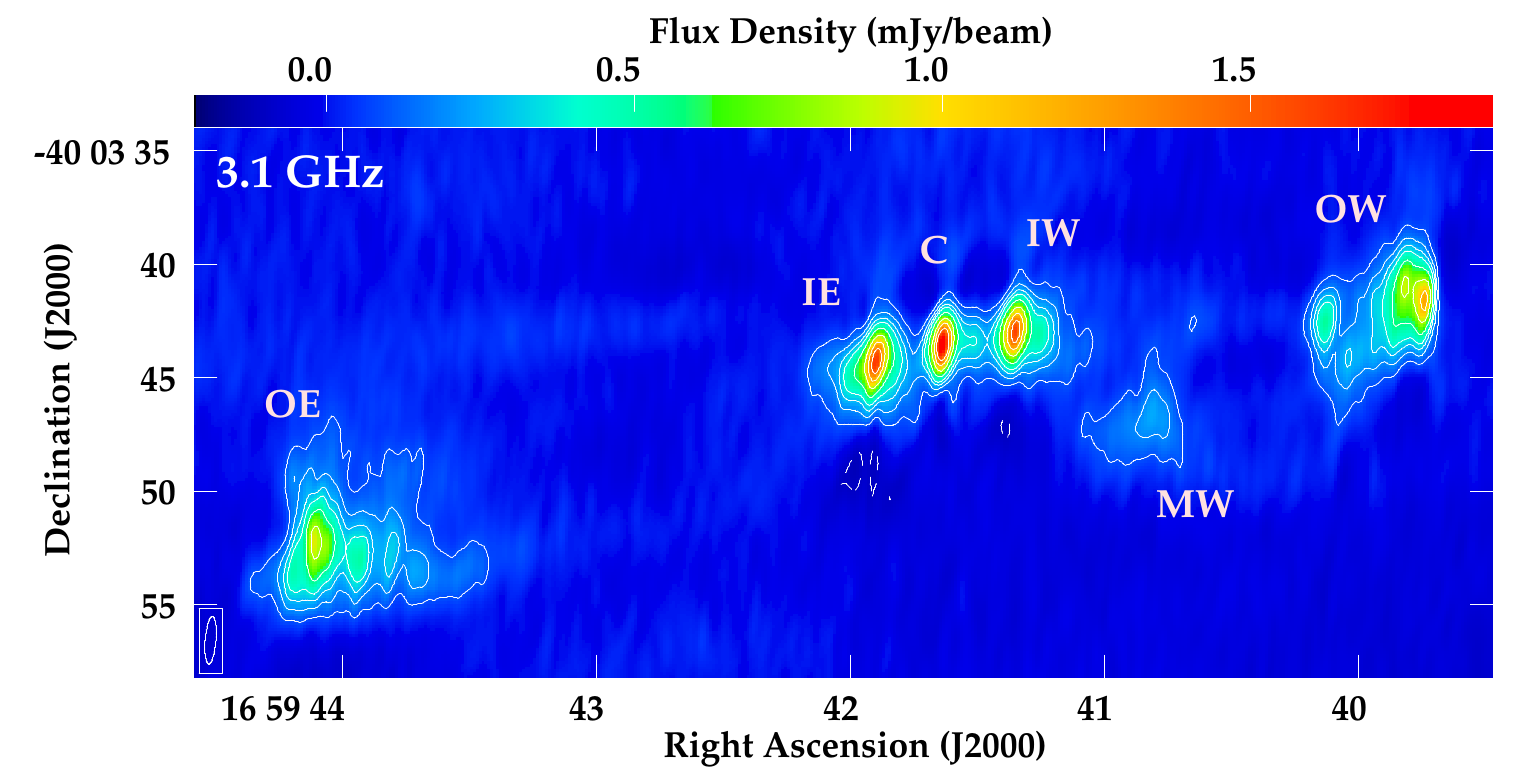}
         \label{fig:sband}
     \end{subfigure}
    \caption{Total intensity (Stokes $I$) images of the radio continuum emission towards I16562 at $6.1$ GHz (top) and $3.1$ GHz (bottom). Labels indicate different continuum components, further expanded upon in Figures \ref{fig:regions_grid_inner} and \ref{fig:regions_grid_outer}. The rms noise for the $6.1$ GHz map is $\sigma = 6.2 \ \mu$Jy/beam, with contours at $\{-5, 5, 10, 20, 40, 80, 160\}\sigma$. The rms noise for the $3.1$ GHz map is $\sigma = 31 \ \mu$Jy/beam, with contours at $\{-3, 3, 6, 9, 12, 18, 27, 36, 45\}\sigma$.}
    \label{fig:main-plot}
\end{figure*}

\section{Results}\label{results}

\subsection{Morphology}\label{morphology}

Figure \ref{fig:main-plot} shows a colormap of the continuum emission detected at 6.1 GHz and 3.1 GHz. 
The overall morphology of I16562 appears as described from the earlier observations of \cite{e_guzman_string_2010} and \cite{e_guzman_2016}, but our higher resolution and sensitivity VLA images reveal a wealth of new details. 
I16562 is comprised of a mix of compact sources and extended regions oriented approximately in the east-west direction with an extent of $\sim 45^{\prime\prime}$, corresponding to $0.35$ pc at the assumed distance of $1.8$ kpc. 
We adopt the nomenclature of \cite{e_guzman_string_2010}, i.e. the previously detected components will be called Central (C), Inner-East (IE), Inner-West (IW), Outer-East (OE), and Outer-West (OW), but we modify the nomenclature to include the detection of a new component (MW), and sub-components of previously detected components.
For the sub-components, we have assigned them numbers from east to west. The position and characteristic of each sub-component are listed in Table \ref{tab:bigtable}. 

We will now discuss the observed components and sub-components in further detail. 
Figures \ref{fig:regions_grid_inner} and \ref{fig:regions_grid_outer} display close-up views of the regions and the sub-components detected at both frequencies. 
In the top two images of Figure \ref{fig:regions_grid_inner}, we show the central and inner sources of the jet at $6.1\,$GHz (left) and $3.1\,$GHz (right). In our nomenclature the peak positions corresponding to the original labeled sources as C, IE and IW by \cite{e_guzman_string_2010} are now labeled C-2, IE-2, and IW-2. 
Component C-2 is the brightest source in the region, and presumably marks the position of the central source driving the jet.
Component IE-2 is separated from Component C-2 by $\sim 3^{\prime\prime}$ eastward, and component IW-2 is $\sim 4^{\prime\prime}$ to the west.

Considering our $6.1\,$GHz image has an angular resolution of $1.03^{\prime\prime}\times 0.26^{\prime\prime}$ at a PA approximately perpendicular to the jet direction, we detect considerable fine structure.
To the west of the central source C-2 we detect two compact sub-components along the axis of the jet which we name C-3, and C-4. To the east of C-2 there is an extension, also along the jet direction. This sub-component we label as C-1.
A similar structure is observed for the IW and IE components. To the east of IE-2 there is an extended source of approximate size $1^{\prime\prime}$ along the jet direction, labeled as source IE-1, and extended emission is also observed to the east of IW-2, labeled as IW-1. A more separated source is detected about $1^{\prime\prime}$ toward the west of IW-2, and is labeled as IW-3. 

Thus, the inner region of the I16562 jet has been resolved in a total of 9 sub-components. 
All of these sources are embedded in extended regions of diffuse emission.
The lower integrated intensity contour at the IE side points along the jet direction east and south to the OE lobe, while on the western side (IW), the lower contour flares out above and below the jet direction to the OW lobe. 
Essentially the same structure is seen in our $3.1\,$GHz image shown in the top right panel of Figure \ref{fig:regions_grid_inner}, albeit at lower angular resolution.

We report the detection of a new component, MW, visible in the lower two panels of Figure \ref{fig:regions_grid_inner}. 
This source lies $\sim 4^{\prime\prime}$ south and $8^{\prime\prime}$ west of the central source C-2. 
At 6.1 GHz, the component is only marginally detected, with peak flux density just above the $5\sigma$ threshold ($\sigma = 6.2\ \mu$Jy/beam), but it is robustly detected at 3.1 GHz, with a peak flux density above $6\sigma = 186\ \mu$Jy/beam. 
The triangular structure of the component appears to be connected to the trailing emission from the southward flare of the jet west of IW-3.

The top panels of Figure \ref{fig:regions_grid_outer} show the structure of component OE at 6.1 and 3.1 GHz, approximately $30''$ east and $10''$ south of the central source C-2. 
We resolve OE into six sub-components distributed over $\sim 7^{\prime\prime}$ along the east–west direction. 
The sub-components are uniformly spaced and aligned along this axis, but are not uniformly compact, with several appearing extended and elongated in the north–south direction. 
The brightest feature, OE-2, shows a compact peak with trailing emission that becomes progressively fainter toward the south. 
Diffuse emission surrounds the feature, extending northward and eastward, and the overall structure appears edge-brightened.
Together, these characteristics give OE the appearance of a chain of knots of different flux densities embedded within a broader envelope of emission.

The bottom panels of Figure \ref{fig:regions_grid_outer} display component OW, located approximately $20''$ west and $3''$ north of the central source C-2. 
We resolve OW into five sub-components extending over $\sim 7^{\prime\prime}$ in the north–south direction and $\sim 4^{\prime\prime}$ along the east–west axis. 
In the eastern half, OW-1 and OW-2 appear as compact sources, faintly connected to a region of more diffuse emission that extends further westward. 
Within this diffuse structure, three additional sub-components, OW-3, OW-4, and OW-5, can be distinguished, with OW-5 being the brightest feature. 
The distribution of the sub-components gives OW a bow-shock–like appearance, with the brightest emission concentrated at its western edge. 
The overall structure appears edge-brightened like OE, but less collimated, with emission spread across a broader area rather than confined to a narrow axis.

In terms of the  larger-scale structure of I16562, we note that
the outer components OE and OW are not equidistant from the Central region. 
OE appears to lie farther from the central emission than OW, and both are offset southward relative to the jet axis as extrapolated from the IE-C-IW position angle. 
As a result, the inner and outer structures are not perfectly aligned. 
Connecting the peak positions of the outer lobes with the Central source, the arms of the ionized jet on either side have a difference of approximately $5^\circ$, subtending an angle of $\sim 175^{\circ}$ between the two arms.

In summary, our images reveal a rich and complex morphology in I16562 with both compact knots and extended structures spanning multiple spatial scales.

\subsection{Flux Densities and Positions}\label{flux_and_posns}

\begin{deluxetable*}{lcccccc}
\tablecaption{Positions and Flux Densities
\label{tab:bigtable}}
\tablehead{
\colhead{Sub-} &
\colhead{Frequency} &
\colhead{RA (J2000)} &
\colhead{Dec (J2000)} &
\colhead{$S_\nu$} &
\colhead{$I_{\nu, \text{peak}}$} &
\colhead{Deconvolved size} \\
\colhead{Component} &
\colhead{(GHz)} &
\colhead{$16^{\mathrm{h}}59^{\mathrm{m}}...$} &
\colhead{$-40\degr03\arcmin...$} &
\colhead{(mJy)} &
\colhead{(mJy beam$^{-1}$)} &
\colhead{$(\theta_{\rm maj}'' \times \theta_{\rm min}'',\,{\rm PA}^\circ)$}
}
\startdata
C-1 & 3.1* & {$41\fs415$} & {$43\farcs16$} & {...} & {$0.63 \pm 0.09$} & {...}\\
{}  & 6.1 & {$41\fs673 \pm 0\fs009$} & {$43\farcs59 \pm 0\farcs01$} & {$1.08 \pm 0.04$} & {$0.35 \pm 0.01$} & {$0.77 \times 0.39, 120.0$}\\
\hline
C-2 & 3.1 & {$41\fs635 \pm 0\fs011$} & {$43\farcs54 \pm 0\farcs03$} & {$3.44 \pm 0.14$} & {$1.88 \pm 0.05$} & {$0.77 \times 0.36, 110.9$}\\
{}  & 6.1 & {$41\fs633 \pm 0\fs009$} & {$43\farcs510 \pm 0\farcs002$} & {$5.13 \pm 0.03$} & {$3.17 \pm 0.01$} & {$0.34 \times 0.20, 111.0$}\\
\hline
C-3 & 3.1* & {$41\fs549$} & {$43\farcs45$} & {...} & {$0.39 \pm 0.09$} & {...}\\
{}  & 6.1 & {$41\fs547 \pm 0\fs017$} & {$43\farcs15 \pm 0\farcs02$} & {$0.19 \pm 0.03$} & {$0.11 \pm 0.01$} & {$0.58 \times 0.23, 20.5$}\\
\hline
C-4 & 3.1 & {$41\fs515 \pm 0\fs114$} & {$43\farcs47 \pm 0\farcs13$} & {$1.54 \pm 0.26$} & {$0.37 \pm 0.05$} & {$1.86 \times 0.71, 89.1$}\\
{}  & 6.1 & {$41\fs502 \pm 0\fs032$} & {$43\farcs15 \pm 0\farcs04$} & {$0.81 \pm 0.06$} & {$0.15 \pm 0.01$} & {$1.23 \times 0.60, 60.4$}\\
\hline
IE-1 & 3.1 & {$42\fs005 \pm 0\fs069$} & {$44\farcs88 \pm 0\farcs14$} & {$1.42 \pm 0.22$} & {$0.42 \pm 0.05$} & {$1.61 \times 1.19, 165.3$}\\
{}   & 6.1 & {$41\fs965 \pm 0\fs018$} & {$44\farcs66 \pm 0\farcs04$} & {$1.11 \pm 0.06$} & {$0.21 \pm 0.01$} & {$1.50 \times 0.67, 155.5$}\\
\hline
IE-2 & 3.1 & {$41\fs897 \pm 0\fs015$} & {$44\farcs30 \pm 0\farcs04$} & {$3.89 \pm 0.17$} & {$1.58 \pm 0.05$} & {$1.41 \times 0.80, 155.5$}\\
{}   & 6.1 & {$41\fs901 \pm 0\fs003$} & {$44\farcs29 \pm 0\farcs01$} & {$2.91 \pm 0.04$} & {$0.91 \pm 0.01$} & {$0.95 \times 0.47, 148.3$}\\
\hline
IW-1 & 3.1* & {$41\fs415$} & {$43\farcs17$} & {...} & {$0.65 \pm 0.09$} & {...}\\
{}   & 6.1 & {$41\fs388 \pm 0\fs005$} & {$43\farcs01 \pm 0\farcs02$} & {$0.80 \pm 0.03$} & {$0.38 \pm 0.01$} & {$0.70 \times 0.38, 175.9$}\\
\hline
IW-2 & 3.1 & {$41\fs351 \pm 0\fs015$} & {$42\farcs98 \pm 0\farcs03$} & {$3.65 \pm 0.17$} & {$1.56 \pm 0.05$} & {$1.09 \times 0.77, 137.8$}\\
{}   & 6.1 & {$41\fs348 \pm 0\fs002$} & {$42\farcs93 \pm 0\farcs01$} & {$2.60 \pm 0.03$} & {$1.16 \pm 0.01$} & {$0.63 \times 0.38, 150.2$}\\
\hline
IW-3 & 3.1 & {$41\fs240 \pm 0\fs065$} & {$43\farcs17 \pm 0\farcs14$} & {$1.49 \pm 0.22$} & {$0.43 \pm 0.05$} & {$1.86 \times 1.14, 170.8$}\\
{}   & 6.1 & {$41\fs253 \pm 0\fs037$} & {$42\farcs90 \pm 0\farcs05$} & {$0.73 \pm 0.06$} & {$0.16 \pm 0.01$} & {$1.72 \times 1.37, 20.5$}\\
\hline
OE-1 & 3.1 & {$44\fs199 \pm 0\fs051$} & {$53\farcs71 \pm 0\farcs15$} & {$1.12 \pm 0.18$} & {$0.40 \pm 0.05$} & {$1.97 \times 0.86, 177.2$}\\
{}   & 6.1 & {$44\fs190 \pm 0\fs018$} & {$53\farcs47 \pm 0\farcs05$} & {$1.01 \pm 0.07$} & {$0.17 \pm 0.01$} & {$2.0 \times 0.67, 172.4$}\\
\hline
OE-2 & 3.1 & {$44\fs105 \pm 0\fs023$} & {$51\farcs84 \pm 0\farcs08$} & {$3.53 \pm 0.22$} & {$0.93 \pm 0.05$} & {$2.84 \times 0.9, 0.16$}\\
{}   & 6.1 & {$44\fs114 \pm 0\fs007$} & {$51\farcs63 \pm 0\farcs03$} & {$1.71 \pm 0.06$} & {$0.33 \pm 0.01$} & {$1.96 \times 0.56, 0.39$}\\
\hline
OE-3 & 3.1 & {$44\fs044 \pm 0\fs027$} & {$52\farcs48 \pm 0\farcs14$} & {$0.65 \pm 0.12$} & {$0.43 \pm 0.05$} & {$1.96 \times 0.18, 1.22$}\\
{}   & 6.1 & {$44\fs045 \pm 0\fs014$} & {$52\farcs33 \pm 0\farcs03$} & {$1.28 \pm 0.06$} & {$0.23 \pm 0.01$} & {$1.63 \times 0.72, 173.5$}\\
\hline
OE-4 & 3.1 & {$43\fs937 \pm 0\fs059$} & {$52\farcs92 \pm 0\farcs13$} & {$2.35 \pm 0.26$} & {$0.50 \pm 0.05$} & {$2.01 \times 1.36, 0.14$}\\
{}   & 6.1 & {$43\fs933 \pm 0\fs015$} & {$53\farcs25 \pm 0\farcs04$} & {$1.00 \pm 0.06$} & {$0.19 \pm 0.01$} & {$1.70 \times 0.65, 178.6$}\\
\hline
OE-5 & 3.1 & {$43\fs810 \pm 0\fs082$} & {$53\farcs08 \pm 0\farcs28$} & {$1.38 \pm 0.26$} & {$0.29 \pm 0.05$} & {$2.97 \times 1.21, 170.8$}\\
{}   & 6.1 & {$43\fs815 \pm 0\fs021$} & {$53\farcs43 \pm 0\farcs07$} & {$0.49 \pm 0.05$} & {$0.11 \pm 0.01$} & {$1.62 \times 0.53, 177.7$}\\
\hline
OE-6 & 3.1 & {$43\fs696 \pm 0\fs150$} & {$53\farcs19 \pm 0\farcs35$} & {$0.62 \pm 0.21$} & {$0.17 \pm 0.05$} & {$1.90 \times 1.37, 31.0$}\\
{}   & 6.1 & {$43\fs711 \pm 0\fs004$} & {$53\farcs49 \pm 0\farcs08$} & {$0.76 \pm 0.08$} & {$0.10 \pm 0.01$} & {$1.67 \times 0.98, 8.39$}\\
\hline
OW-1 & 3.1 & {$40\fs129 \pm 0\fs025$} & {$42\farcs61 \pm 0\farcs07$} & {$1.46 \pm 0.11$} & {$0.52 \pm 0.03$} & {$1.30 \times 0.92, 174.2$}\\
{}   & 6.1 & {$40\fs128 \pm 0\fs009$} & {$42\farcs50 \pm 0\farcs02$} & {$0.92 \pm 0.04$} & {$0.24 \pm 0.01$} & {$1.05 \times 0.63, 174.0$}\\
\hline
OW-2 & 3.1 & {$40\fs031 \pm 0\fs012$} & {$44\farcs03 \pm 0\farcs23$} & {$1.77 \pm 0.30$} & {$0.28 \pm 0.01$} & {$2.72 \times 1.78 , 167.4$}\\
{}   & 6.1 & {$40\fs038 \pm 0\fs002$} & {$44\farcs11 \pm 0\farcs04$} & {$0.60 \pm 0.04$} & {$0.13 \pm 0.01$} & {$1.32 \times 0.71, 0.29$}\\
\hline
OW-3 & 3.1* & {$39\fs844$} & {$42\farcs56$} & {...} & {$0.57 \pm 0.09$} & {...}\\
{}   & 6.1 & {$39\fs846 \pm 0\fs026$} & {$42\farcs79 \pm 0\farcs04$} & {$1.07 \pm 0.06$} & {$0.14 \pm 0.01$} & {$1.51 \times 1.10, 169.2$}\\
\hline
OW-4 & 3.1 & {$39\fs813 \pm 0\fs022$} & {$41\farcs07 \pm 0\farcs08$} & {$3.22 \pm 0.16$} & {$0.88 \pm 0.02$} & {$2.44 \times 1.26, 170.0$}\\
{}   & 6.1 & {$39\fs816 \pm 0\fs009$} & {$40\farcs83 \pm 0\farcs02$} & {$2.07 \pm 0.06$} & {$0.31 \pm 0.01$} & {$1.71 \times 0.84, 168.4$}\\
\hline
OW-5 & 3.1 & {$39\fs753 \pm 0\fs011$} & {$41\farcs54 \pm 0\farcs03$} & {$3.48 \pm 0.24$} & {$1.17 \pm 0.03$} & {$1.81 \times 0.95, 4.01$}\\
{}   & 6.1 & {$39\fs739 \pm 0\fs002$} & {$41\farcs44 \pm 0\farcs01$} & {$1.76 \pm 0.03$} & {$0.64 \pm 0.01$} & {$1.13 \times 0.39, 4.91$}\\
\hline
MW* & 3.1 & {$40\fs815$} & {$46\farcs59$} & {$1.72 \pm 0.03$} & {$0.26 \pm 0.09$} & {...}\\
{}   & 6.1 & {$40\fs750$} & {$46\farcs70$} & {$1.05 \pm 0.01$} & {$0.07 \pm 0.02$} & {...}\\
\enddata
\tablenotetext{*}{The reported RA and Dec for MW and the starred $3.1$~GHz sub-components are the position of its peak intensity. The flux density reported for MW comes from the \texttt{\textcolor{blue}{TVSTAT}} command in AIPS by tracing the lowest level of detection contour and combining the intensity of each pixel within the traced area.}
\end{deluxetable*}

We report the positions and flux densities of each identified sub-component at 3.1 and 6.1 GHz in Table \ref{tab:bigtable}, together with the peak intensity and deconvolved source size parameters derived using the interactive Astronomical Image Processing System\footnote{\url{https://www.aips.nrao.edu/}} (AIPS) task \codeword{TVSAD}.

Accurate astrometry is critical for reliable position and flux measurements. 
Our 6.1 GHz phase calibrator, J1744$–$3116, exhibits extended structure at this frequency and is partially resolved on short baselines, potentially compromising absolute astrometric accuracy. 
In contrast, the 3.1 GHz calibrator, J1626$-$2951, has excellent astrometric accuracy ($<2$ mas) and behaves as a point source at centimeter wavelengths. 
We therefore adopt the 3.1 GHz image as the astrometric reference frame and align the peak intensity pixel between the two frequency bands by applying a uniform coordinate shift of $-0\fs010$ and $-0\farcs14$ in the RA and Declination axes, respectively, to the 6.1 GHz image. 
All positions reported in Table \ref{tab:bigtable} include this astrometric correction.

The \codeword{TVSAD} task identifies candidate sources in the image and fits two-dimensional Gaussian models to each detection. 
Positional and flux density uncertainties derived from \codeword{TVSAD} were verified using the \codeword{imfit} task in CASA. 
At 6.1 GHz, all sources listed in Table \ref{tab:bigtable} were successfully modeled with two-dimensional Gaussians. 
At 3.1 GHz, however, several sources appear blended or merged with nearby emission, preventing reliable decomposition into individual components. 
In such cases, Gaussian fitting algorithms in both AIPS and CASA were unable to uniquely identify multiple peaks. 
For these sources, we therefore report only the peak intensity, $I_{\nu, \text{peak}}$, at their~$6.1$~GHz positions rather than integrated flux densities or fitted positions. 
We also note that since these sources are blended with other continuum sources, the peak intensity is an upper limit. 
The uncertainty in $I_{\nu, \text{peak}}$ for these specific sources is reported to be the $3\sigma$ rms noise values of the image ($\sim 93 \ \mu \text{Jy beam}^{-1}$). 

\begin{figure*}[!tbp]
    \centering
    \begin{subfigure}[t]{0.482\textwidth}
        \centering
        \includegraphics[width=\textwidth]{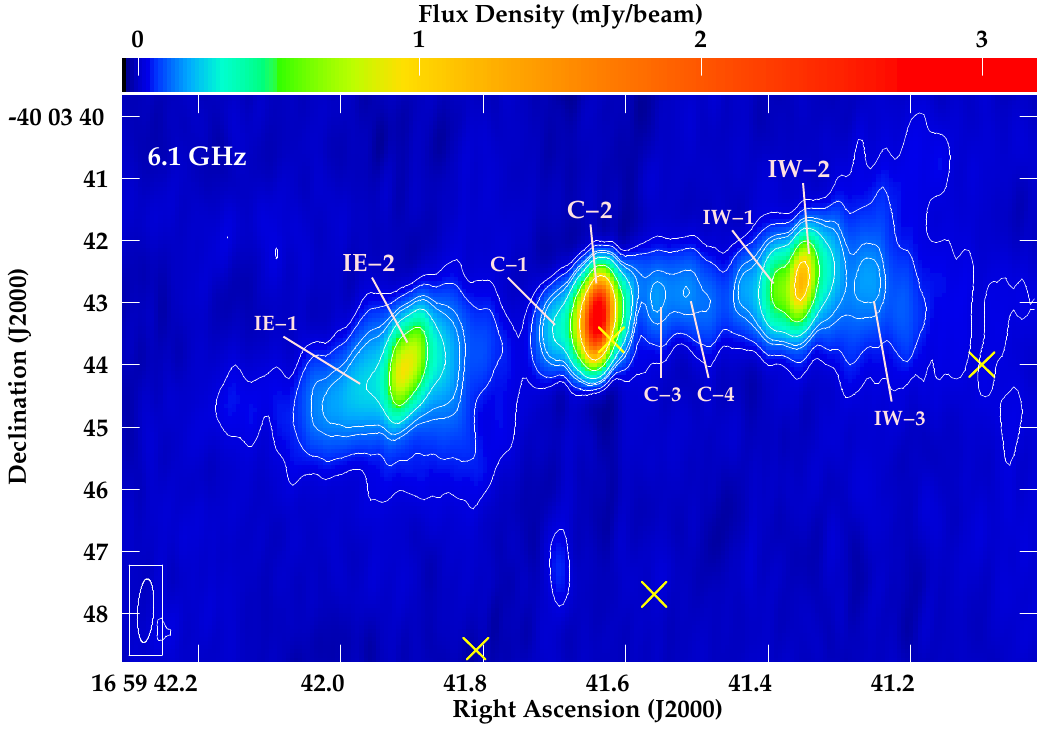}
        \label{fig:central_c}
    \end{subfigure}
    \hspace{1em}
    \begin{subfigure}[t]{0.482\textwidth}
        \centering
        \includegraphics[width=\textwidth]{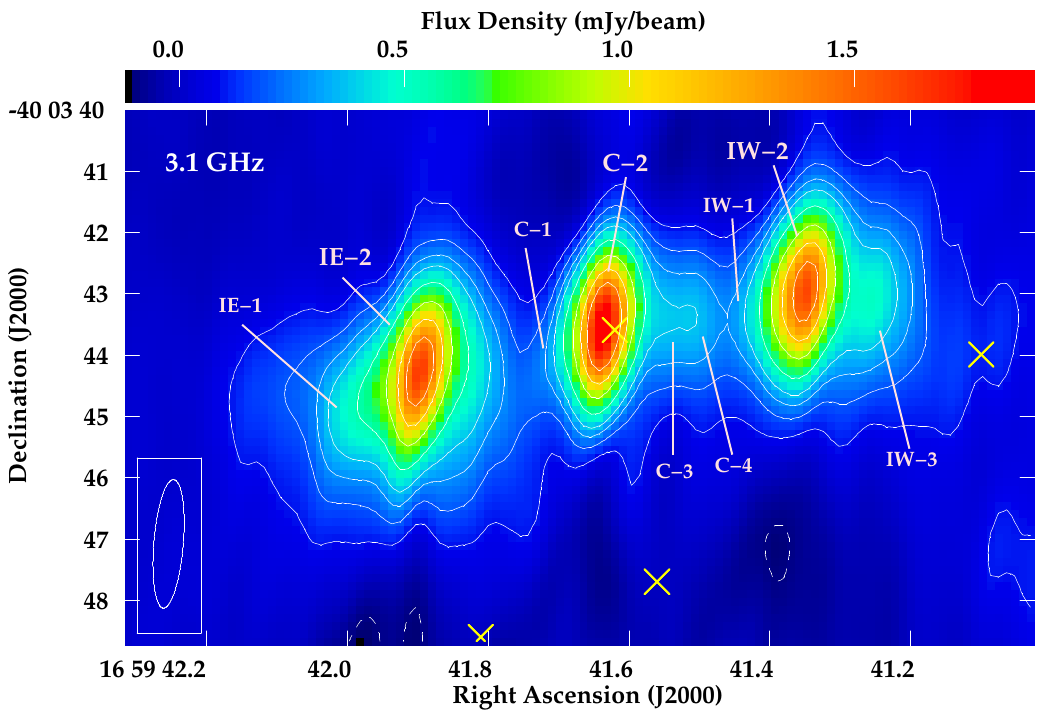}
        \label{fig:central_s}
    \end{subfigure}

    \vspace{1em}
    
    \begin{subfigure}[t]{0.482\textwidth}
        \centering
        \includegraphics[width=\textwidth]{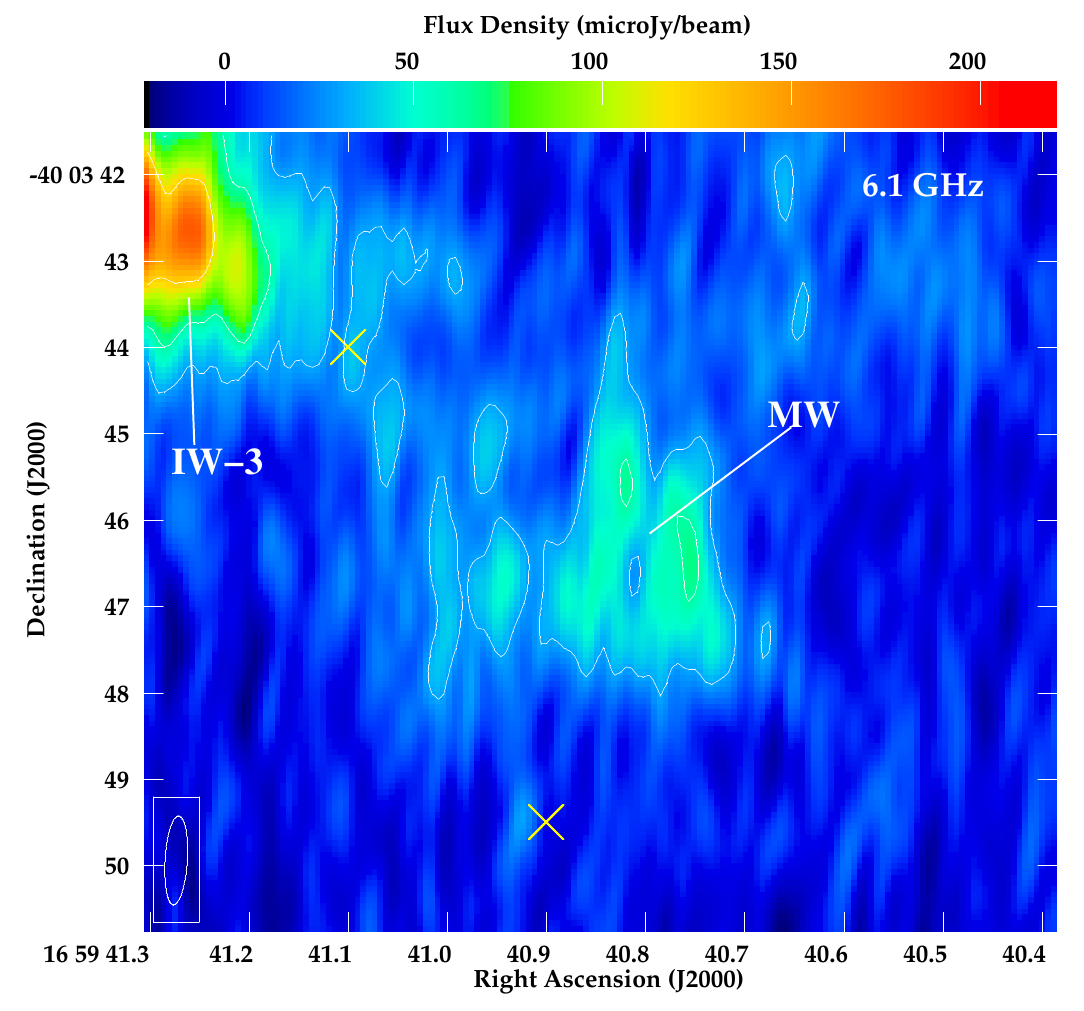}
        \label{fig:mw_c}
    \end{subfigure}
    \hspace{1em}
    \begin{subfigure}[t]{0.482\textwidth}
        \centering
        \includegraphics[width=\textwidth]{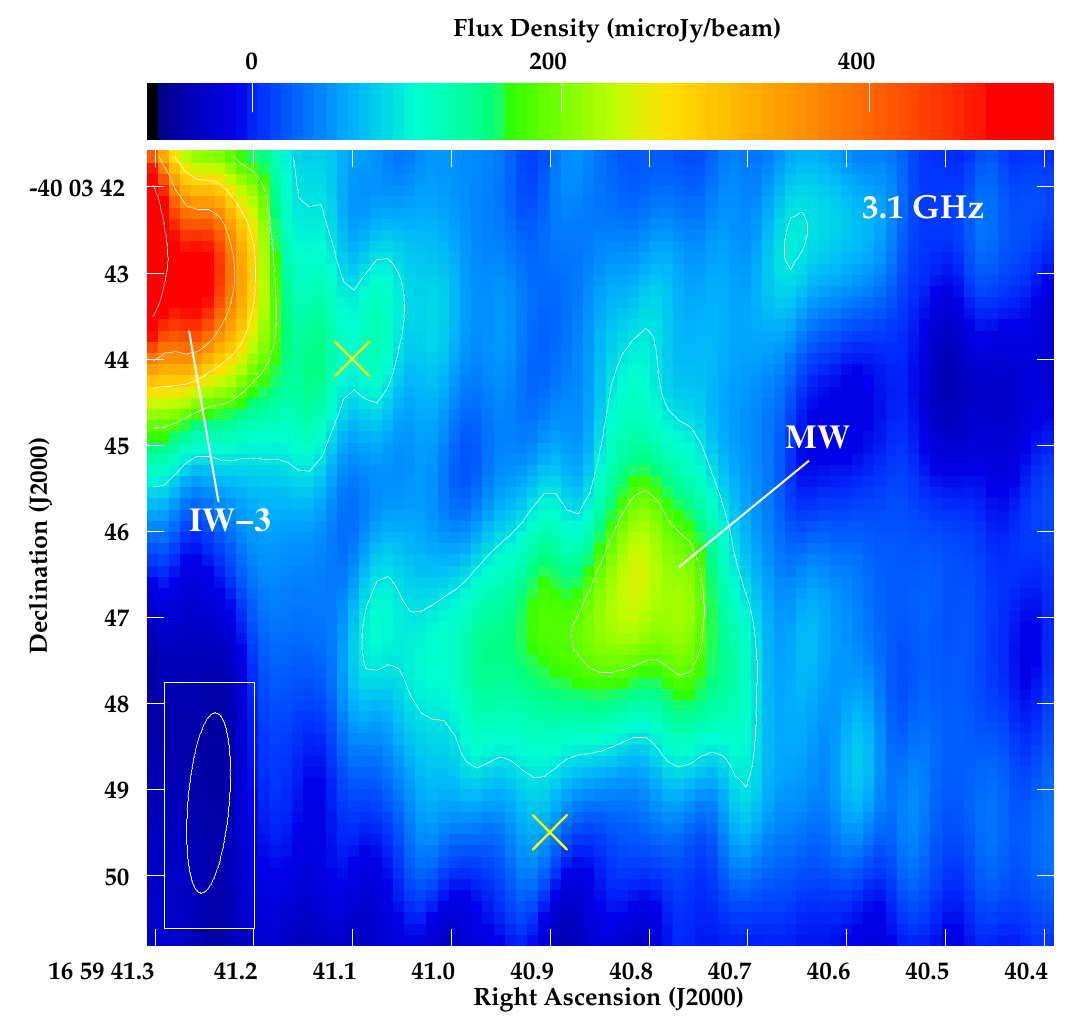}
        \label{fig:mw_s}
    \end{subfigure}

    \caption{Continuum emission cutouts for the Central (top) and MW (bottom) regions of I16562, at $6.1$ GHz (left) and $3.1$ GHz (right). Yellow crosses on the maps denote Chandra X-ray sources reported in \cite{montes_chandra_2020}. The rms noise for the $6.1$ GHz map is $\sigma = 6\ \mu$Jy/beam, with contours at $\{-5, 5, 10, 20, 25, 40, 80, 160\}\sigma$. The rms noise for the $3.1$ GHz map is $\sigma = 31 \ \mu$Jy/beam, with contours at $\{-3, 3, 6, 9, 12, 18, 27, 36, 45\}\sigma$.}
    \label{fig:regions_grid_inner}
\end{figure*}

\begin{figure*}[!tbp]
    \centering
    \begin{subfigure}[t]{0.482\textwidth}
        \centering
        \includegraphics[width=\textwidth]{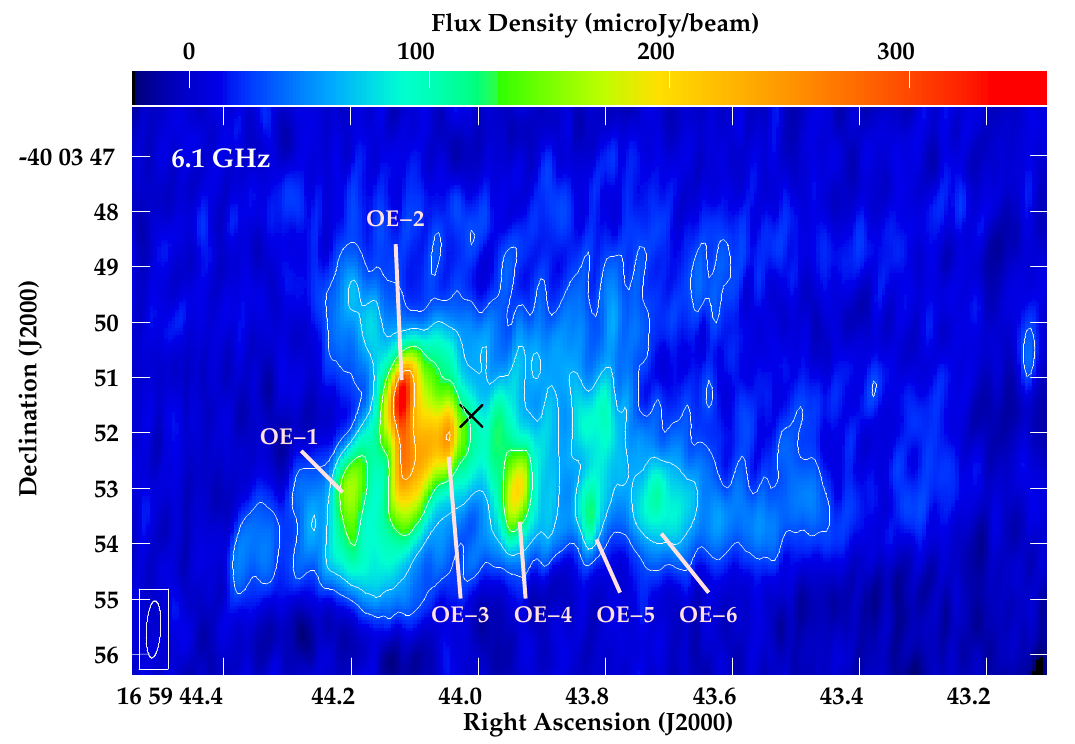}
        \label{fig:oe_c}
    \end{subfigure}
    \hspace{1em}
    \begin{subfigure}[t]{0.482\textwidth}
        \centering
        \includegraphics[width=\textwidth]{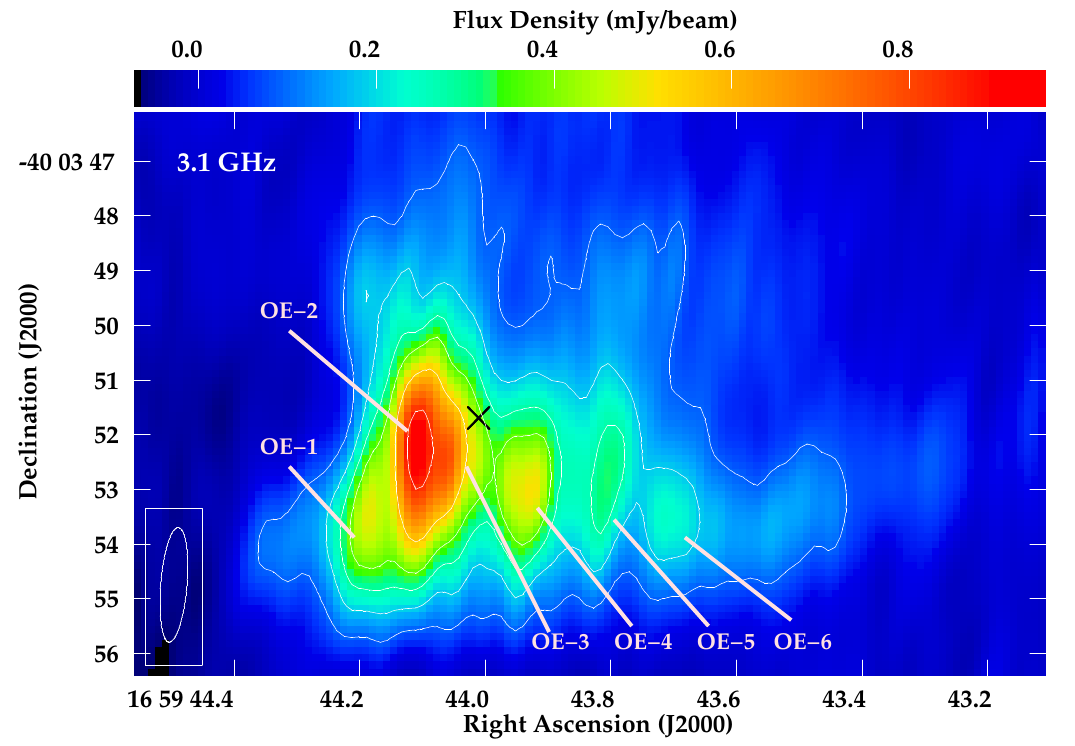}
        \label{fig:oe_s}
    \end{subfigure}

    \vspace{1em}
    
    \begin{subfigure}[t]{0.482\textwidth}
        \centering
        \includegraphics[width=\textwidth]{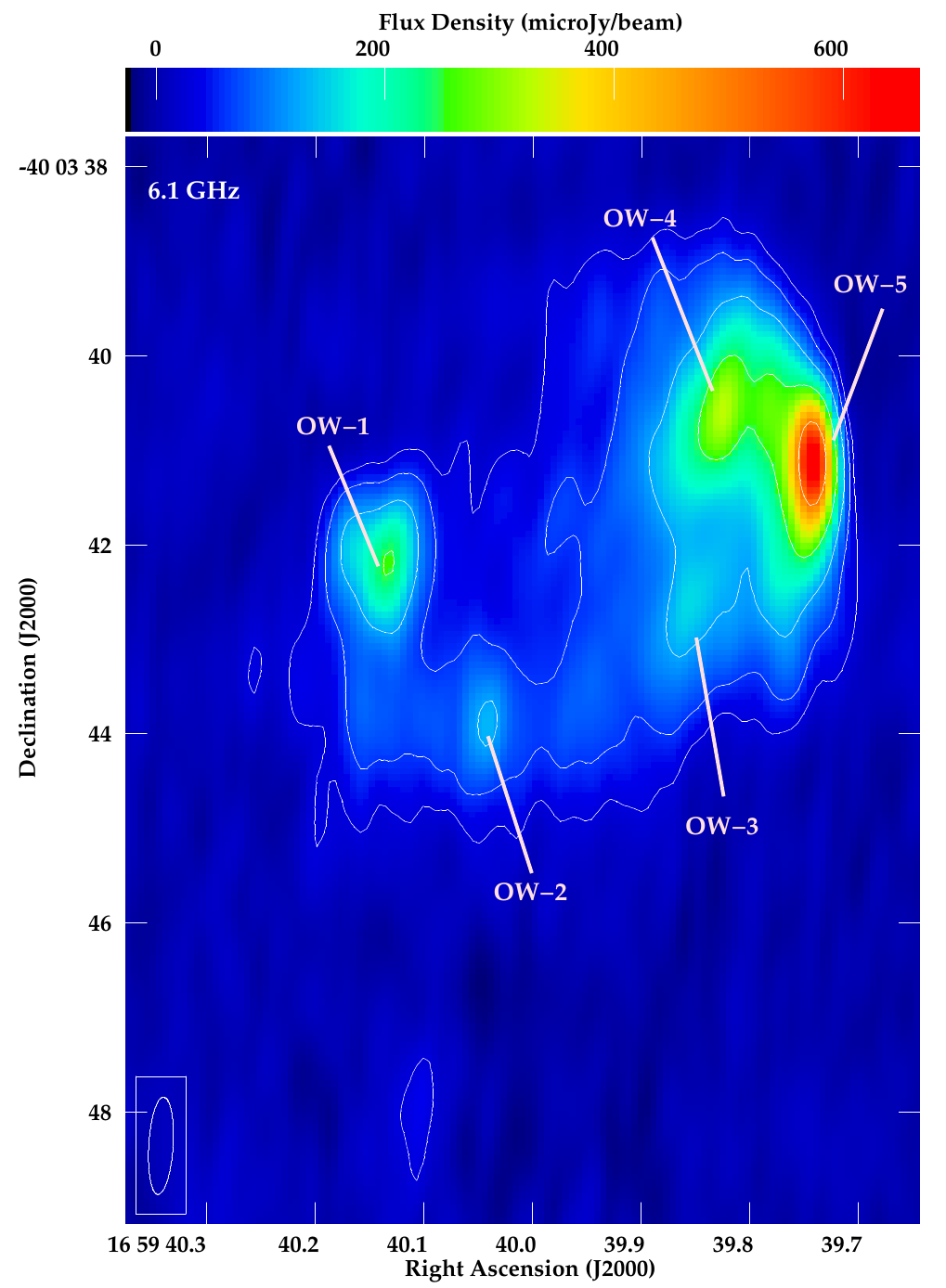}
        \label{fig:ow_c}
    \end{subfigure}
    \hspace{1em}
    \begin{subfigure}[t]{0.482\textwidth}
        \centering
        \includegraphics[width=\textwidth]{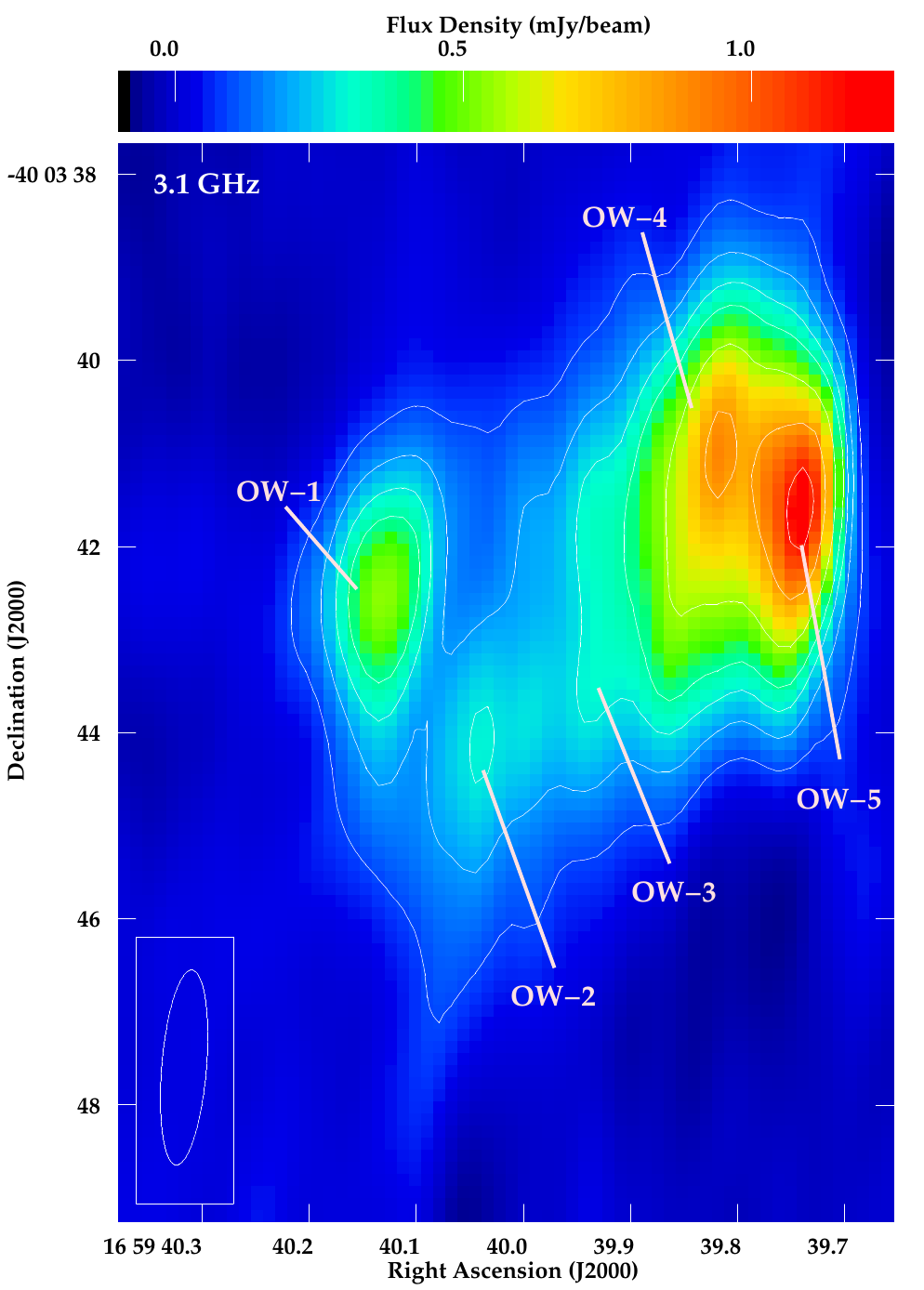}
        \label{fig:ow_s}
    \end{subfigure}

    \caption{Continuum emission cutouts for the outer regions OE (top) and OW (bottom) of I16562, at $6.1$ GHz (left) and $3.1$ GHz (right). Black crosses on the maps denote Chandra X-ray sources observed in \cite{montes_chandra_2020}. Contours used in the cutouts are the same as in Figure \ref{fig:regions_grid_inner}.}
    \label{fig:regions_grid_outer}
\end{figure*}

As noted below Table \ref{tab:bigtable}, the position reported for source MW corresponds to the location of its peak brightness. 
Source MW, shown in the bottom panel of Figure \ref{fig:regions_grid_inner}, exhibits highly non-Gaussian and asymmetric morphology at both frequency bands, precluding reliable Gaussian fitting. 
Instead, we defined regions encompassing emission above $31 \ \mu\text{Jy beam}^{-1}$ at 6.1 GHz and above $93 \ \mu\text{Jy beam}^{-1}$ at 3.1 GHz and used the AIPS task \codeword{TVSTAT} to measure the peak brightness, integrated flux density, and corresponding position. 
The total area of each region was recorded, and rms noise levels were estimated from multiple source-free regions of identical area across the image to determine the flux density uncertainty. 
The uncertainty in peak brightness for source MW was also calculated using the $3\sigma$ rms noise values.

We also identify four compact radio sources within the field of I16562. 
Sources 1 - 3 were previously reported in \citet{e_guzman_2016} at $9, 17, \ \text{and} \ 19$ GHz, and a new source we name Source 4.
Sources~1--3 are detected at 6.1~GHz but not at 3.1~GHz, whereas Source~4 is detected in both bands.
Since these sources were not observed at 5~GHz previously by \citet{e_guzman_2016}, we do not have an equivalent observation to compare our values with, and we report flux and position measurements for these sources using our 6.1~GHz map.

Source 1 is detected at $\alpha_\mathrm{J2000} = 16^{\mathrm{h}}59^{\mathrm{m}}42\fs001$, $\delta_\mathrm{J2000} = -40\degr03\arcmin31\farcs18$ with $S_\nu = 67.3 \pm 3.1~\mu$Jy, and is unresolved. 
\citet{e_guzman_2016} infer that this source may be extragalactic in nature, with a potential galaxy candidate about $\sim4\farcs8$ away. 
We report Source 2 to be at $\alpha_\mathrm{J2000} = 16^{\mathrm{h}}59^{\mathrm{m}}42\fs862$, $\delta_\mathrm{J2000} = -40\degr03\arcmin35\farcs79$, with a flux density of $S_\nu = 72.5 \pm 7.2~\mu$Jy. 
Coincident with a GLIMPSE object \citep{e_guzman_2016}, a millimeter continuum peak \citep{Guzman2014}, and a 6.7~GHz Class~II methanol maser \citep{Caswell2010}, Source 2 likely traces an embedded high-mass protostar. 
Source 3, at $\alpha_\mathrm{J2000} = 16^{\mathrm{h}}59^{\mathrm{m}}41\fs693$, $\delta_\mathrm{J2000} = -40\degr03\arcmin47\farcs28$, with a flux of $S_\nu = 92.3 \pm 5.6~\mu$Jy, is marginally resolved. 
\citet{e_guzman_2016} believes that due to the rising spectrum that they observe at 9, 17, and 19~GHz, the likely origin of this source is free-free emission from a low-mass YSO.  

Finally, Source 4 is a newly detected source located $44\farcs8$ northwest from the jet ($\sim7.6\times10^{4}$~au, or 0.37~pc) at $\alpha_\mathrm{J2000} = 16^{\mathrm{h}}59^{\mathrm{m}}41\fs135$, $\delta_\mathrm{J2000} = -40\degr02\arcmin59\farcs10$ with $S_{6 \ \text{GHz}}= 97.0 \pm 4.8~\mu$Jy and $S_{3 \ \text{GHz}}= 169.0 \pm 13.0~\mu$Jy. The resulting
spectral index after matched-beam imaging is $\alpha = -0.39 \pm 0.16$, indicating non-thermal emission. 
This source has no object associated with it within $5^{\prime\prime}$ in the SIMBAD or GLIMPSE databases, but data from \citet{montes_chandra_2020} shows that there is an X-ray source with a flux of $F_x = 0.33 \times 10^{-14} \ \text{erg s}^{-1}\text{cm}^{-2}$ coincident with Source~4. 
\citet{Rosero2019} has shown that at an rms noise of $\sim 11 \ \mu\text{Jy beam}^{-1}$ and within a distance of $\sim 2$ kpc, it is possible to detect T-Tauri stars in our Galaxy at 6.2~GHz using the VLA, with a coronal (gyrosynchrotron) origin. 
Taking both of these findings into account, Source 4 could be a T-Tauri star.

\subsection{Spectral Index}\label{spindex}

\begin{figure*}[!tbp]
    \centering
    \includegraphics[width=1\linewidth]{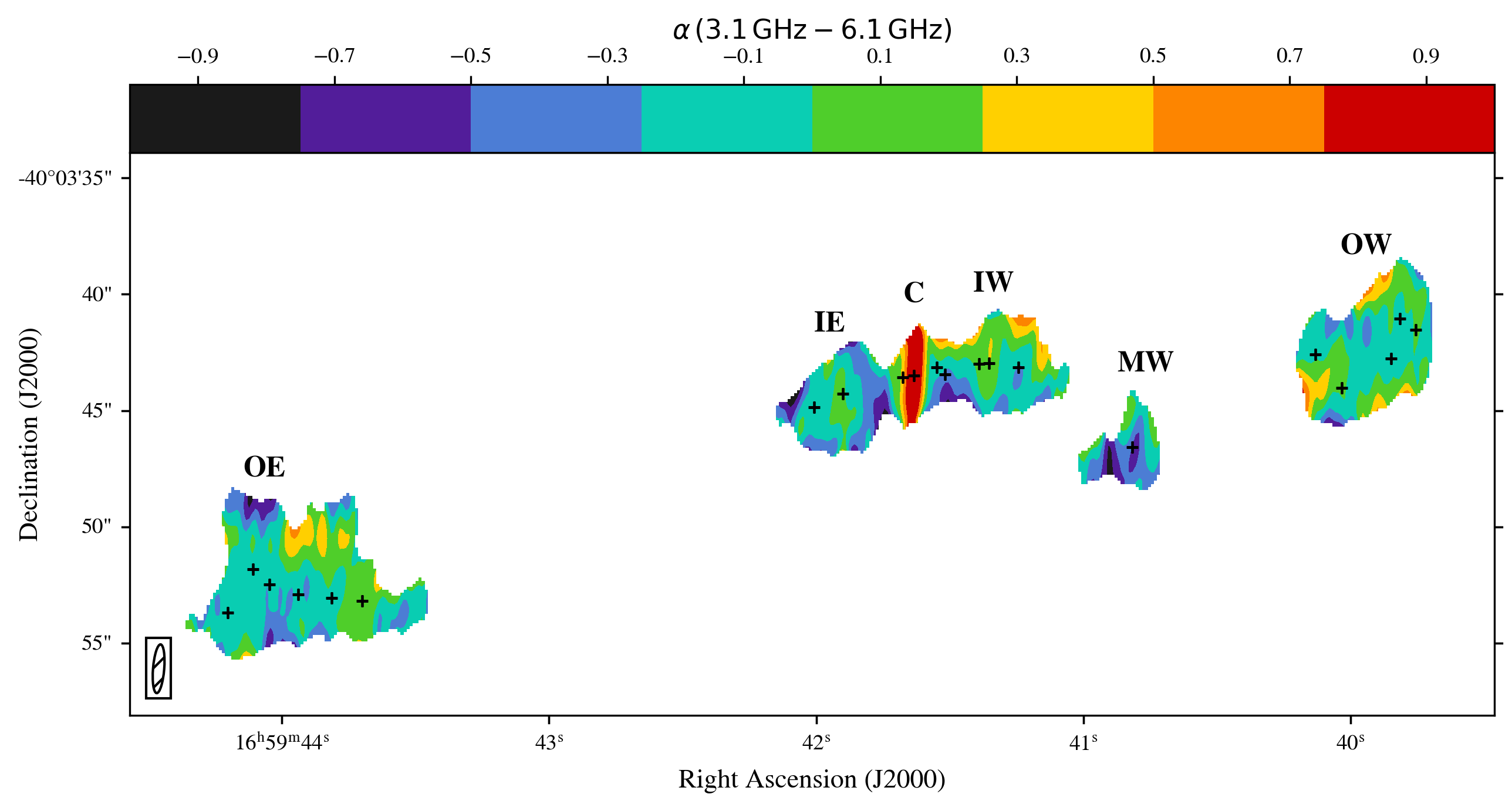}
    \caption{Spectral Index $(\alpha)$ mapping of the emission at 3.1~GHz vs 6.1~GHz. The beam size is the same as that of the 3.1~GHz image in Figure \ref{fig:main-plot}. The map has been blanked for all pixels below $95 \ \mu\text{Jy beam}^{-1} \ (\sim 3\sigma_\text{3.1 GHz})$, in order to only display the spectral index of regions with significant emission. The crosses (+) represent the fitted positions of all sub-components in each lobe at $6.1$~GHz.}
    \label{fig:alpha}
\end{figure*}

Due to the complexity of the individual sub-components and the difficulty to clearly distinguish them in the low frequency band, we do not report the spectral slope of each sub-component in Table \ref{tab:bigtable}. 
Instead, in order to understand the behavior of both the compact components and the diffuse emission with respect to wavelength, we present the spectral index as a continuous map in Figure \ref{fig:alpha}. 

We convolved the 6.1~GHz image to the same synthesized beam size as the 3.1~GHz image, and then smoothed and re-gridded the pixels to match the same cell size as that of the 3.1~GHz image. 
The AIPS command \codeword{COMB} with parameter \codeword{OPCODE} = \codeword{SPIX} was used next in order to take the two matching images and map the spectral index of each pixel. 
All pixels below $95 \ \mu\text{Jy beam}^{-1}$ ($\sim 3 \sigma_\text{3.1 GHz}$) were then blanked in both images so that the spectral index would only be mapped for regions with significant emission. 

Figure \ref{fig:alpha} reaffirms results of previous investigations of I16562, and also shows some interesting results for the newly detected sources. 
Similar to what \citet{e_guzman_2016} reports, the central source, corresponding to C-2 in our nomenclature, has a spectral index $\alpha \geq 1$, representative of thermal emission. 
IE, IW, and the outer regions have a relatively flat spectral index in the range $-0.1 < \alpha < 0.1$, which is consistent with optically thin free-free emission. 

There are regions of negative spectral index along the edges of IE and OE, and near to some of the sub-components which could be due to non-thermal emission. 
However, due to their small dimensions and location, they may be spurious. In contrast, for the newly detected source MW the spectral index $\alpha  \leq - 0.6$ for a significant amount of the encompassed region, providing strong evidence that source MW exhibits non-thermal emission. 

\section{Analysis}\label{analysis}

\subsection{Variability \& Comparison with Previous Maps}\label{variability}
As discussed above, variability in ionized jets points to episodic accretion events, and hence one of the goals of this work was to investigate the variability of the radio sources within I16562. 
By comparing flux densities across multiple epochs spanning more than eight years \citep[this work]{e_guzman_string_2010, e_guzman_2016}, we can test whether flux density variability of both previously known, and newly identified components has occurred.

We derived 5~GHz flux densities for the compact and extended components in I16562 from our 2017 VLA C-band observations. 
To ensure a direct comparison with the 2014/15 ATCA data presented by \citet{e_guzman_2016}, we restricted the MTMFS image to a frequency range of 4--6~GHz, corresponding to an effective central frequency of 5~GHz. 
The resulting image, which had a synthesized beam of $\theta_\text{FWHM} = 1\farcs03\times0\farcs26$,~PA~=~$-3\fdg43$, was convolved to match the ATCA 5~GHz synthesized beam ($\theta_\text{FWHM} = 2\farcs49\times1\farcs36$,~PA~=~$3\fdg00$).

Flux densities were measured following the same approach as in \citet{e_guzman_2016}. 
For the inner components (C, IE, and IW), two-dimensional Gaussian fits were performed using the \codeword{imfit} task in CASA. 
For the outer components (OE and OW) the total emission was integrated within $9''\times9''$ boxes centered on each lobe. 

When comparing our flux measurements with previous epochs we need to keep in mind the associated uncertainties and systematic differences. Because the observations were made with different instruments (VLA vs ATCA),
different calibrators, $uv$-coverage, instrumental setups, and atmospheric conditions between epochs, we
consider the comparison between the tabulated flux densities as approximate.
In particular, the measured flux densities for the several arcsec sized OE and OW regions could be affected by different $uv$-coverage,
resulting in less sensitivity to extended emission and hence lower flux densities in our higher angular resolution data. This effect is exacerbated by map artifacts in our VLA data which were amplified by the convolution to the ATCA beam.

In Table~\ref{tab:flux_comparison} we compare our flux densities with those reported in previous epochs. 
The comparison reveals distinct trends among the inner and outer components. 
The inner components C, IE, and IW exhibit decreases in flux of approximately 25--30\%, and we note that 
this decrease correlates with the increasing angular resolution of successive observations. In particular, the central source C exhibits the largest relative drop, from 8.1~mJy in 2008/09 to 5.9~mJy in 2017. 
Given that higher-resolution data are more sensitive to compact emission but less so to extended structure, these trends are best explained by differences in resolution and $uv$-coverage rather than intrinsic variability. 
The eastern and western outer lobes show divergent behavior: OE remains nearly constant within uncertainties, whereas OW shows an apparent increase in brightness.   
Its measured flux density rises steadily from 6.2~mJy in 2008/09 to 7.1~mJy in 2014/15, and to 8.1~mJy in our 2017 observations. 
This monotonic increase, consistent over nearly a decade, suggests genuine variability rather than a calibration or imaging artifact. However, considering the uncertainties, this is still a marginal result that must be confirmed by future observations.  

\begin{figure}
    \centering
    \includegraphics[width=1\linewidth]{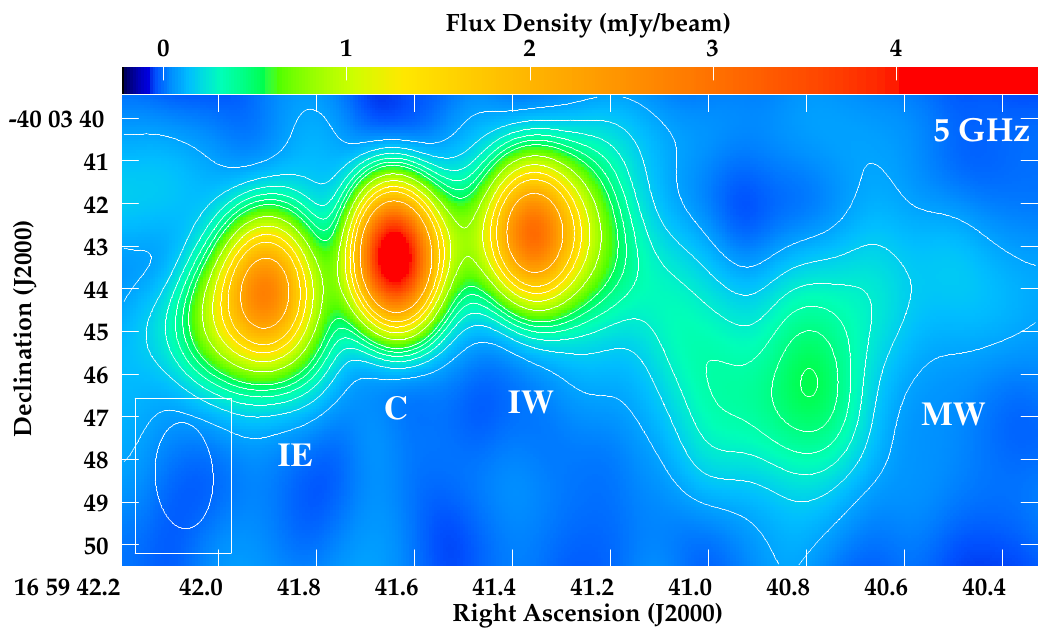}
    \caption{5~GHz smoothed continuum map of the inner components - IE, C, IW, MW from east to west constructed using our VLA data. Synthesized beam parameters chosen using \citet{e_guzman_2016} of $\theta_\text{FWHM} = 2\farcs49\times1\farcs36$,~PA~=~$3^\circ$. The rms noise in this image is $\sigma = 30 \ \mu\text{Jy beam}^{-1}$, with contour levels at $\{-3, 3, 6, 9, 12, 15, 18, 27, 36, 45, 54, 72\}\sigma$.}
    \label{fig:mw_5ghz_smoothed}
\end{figure}

\begin{figure}
    \centering
    \includegraphics[width=1\linewidth]{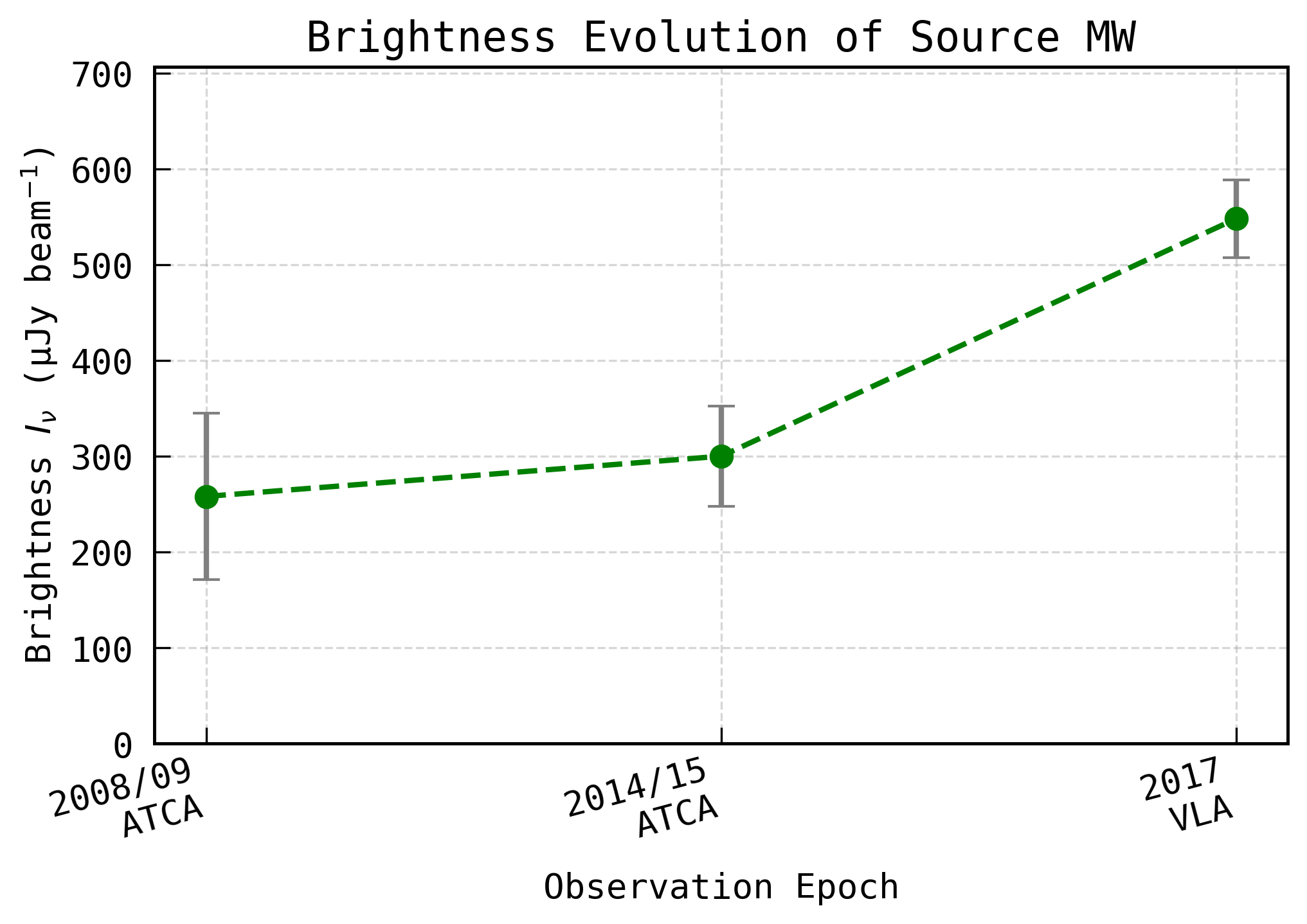}
    \caption{Comparison of the peak brightness in Source MW detected across three epochs of observations. Note that the brightness of the 2008/09 and 2014/15 ATCA data is an estimation from the highest contour level present in the region of component MW in the continuum plots presented in \citet{e_guzman_string_2010} and \citet{e_guzman_2016}. The uncertainties for the brightness were calculated using $\sigma_\text{unc} = \sqrt{\sigma_\text{rms}^2 + (f_\text{cal} \ I)^2}$, where $f_\text{cal}$ is a fractional calibration uncertainty term ($\sim 5\%$), and $I$ is the brightness.}
    \label{fig:mw_brightness}
\end{figure}

The newly identified source MW exhibits evidence of significant variability in our data. 
In the 2008–2009 ATCA data of \citet{e_guzman_string_2010}, it appears only at the lowest level of significance ($\sim 3 \sigma$, corresponding to $\sim 258~ \mu\text{Jy}\ \text{beam}^{-1}$) in the 4.8 GHz map. 
In the later observations of \citet{e_guzman_2016}, the same feature is marginally detected at a peak level of $\sim 6\sigma$ ($\sim 300~ \mu\text{Jy}\ \text{beam}^{-1}$) in the 5 GHz map, and was not classified as a distinct component. 
In contrast, our 5 GHz image — smoothed to match the \citet{e_guzman_2016} beam — has a peak brightness of $ 548 ~\mu \text{Jy} \ \text{beam}^{-1}$, corresponding to a detection significance exceeding 18$\sigma$ ($\sigma \sim 30 ~\mu \text{Jy} \ \text{beam}^{-1}$), i.e. the intensity appears to have nearly doubled compared to the earlier observations.
This represents a substantial increase in intensity relative to previous epochs, indicating positive variability in MW.
Figure \ref{fig:mw_5ghz_smoothed} displays the the smoothed 5~GHz image of source MW. In this image, a connection to the western jet is also clearly visible.
Figure \ref{fig:mw_brightness} shows the evolution of the maximum brightness of Source MW across the three epochs.

As discussed in the previous section, we also detect four point sources in our $6\,$GHz map. 
Sources 1, 2 and 3 were reported by Guzm\'an et al. (2016) at $9, 17$ and $19\,$GHz, but not detected in their $5\,$GHz map. 
Given the rms noise value of $\sigma = 18\,\mu$Jy$\,$beam$^{-1}$ of their $5\,$GHz map, if these sources were constant at the level we report in Table \ref{tab:flux_comparison}, they should have appeared in their $5\,$GHz map with significance levels $3.7, 4.1$ and $5.1\,\sigma$, respectively. 
Similarly, the newly detected source 4, which is located $\approx 45^{\prime\prime}$ to the north of the ionized jet, would have appeared at $5.4\,\sigma$. 
Hence variability for these sources is a distinct possibility.

\begin{deluxetable}{lccc}[!h]
    \tabletypesize{\footnotesize}
    \tablewidth{0pt}
    \setlength{\tabcolsep}{3pt}
    \tablecaption{Flux Densities at 5~GHz Across Epochs \label{tab:flux_comparison}}
    \tablehead{
        \colhead{Source} &
        \colhead{2017 VLA} &
        \colhead{2014/15 ATCA} &
        \colhead{2008/09 ATCA\tablenotemark{a}}
    }
    \startdata
        C  & $5.91 \pm 0.72$ & $7.6 \pm 0.2$ & $8.1 \pm 0.2$ \\
        IE & $4.74 \pm 0.20$ & $6.0 \pm 0.3$ & $6.3 \pm 0.2$ \\
        IW & $5.10 \pm 0.29$ & $6.0 \pm 0.3$ & $5.8 \pm 0.2$ \\
        OE & $8.64 \pm 0.70$ & $8.6 \pm 0.1$ & $9.0 \pm 0.3$ \\
        OW & $8.08 \pm 0.70$ & $7.1 \pm 0.1$ & $6.2 \pm 0.2$ \\
    \enddata
    \tablecomments{Flux densities in mJy. \tablenotemark{a}~4.8~GHz.}
\end{deluxetable}

\subsection{Proper Motion}\label{proper_motion}

I16562 has been observed across multiple epochs, providing a basis for investigating the proper motions (PMs) of the jet lobes associated with it. 
\citet{e_guzman_2016} report PM measurements at 5 and 9~GHz derived from observations spanning a timescale of $\sim\,$6~years (2008/09 -- 2014/15). 
At 9~GHz, they determine PM magnitudes in the range $17$--$64$~mas~yr$^{-1}$, with the lobes receding from the central source along the jet axis. 
Our observations were carried out in 2017 January. 
Adopting a mean epoch of 2015 February for the \citet{e_guzman_2016} 2014/15 observations, the time baseline between the two datasets is $\Delta t \approx 1.92$~yrs. 

To enable a cross-epoch comparison, we convolved our 6.1~GHz image to match the resolution of the \citet{e_guzman_2016} images, resampled to a common pixel scale of $0\farcs5$~pixel$^{-1}$, and aligned the coordinate frames by matching the peak brightness of source C-2 to the position of source C reported by \citet{e_guzman_2016}. 
Lobe positions were then measured following similar methods as outlined in their paper: \codeword{imfit} for the inner lobes and peak brightness identification via \codeword{imstat} for the outer lobes in CASA, with position uncertainties derived following \citet{condon_errors_1997}.

In all cases the measured displacements of the lobes did not reach a level of statistical significance sufficient to claim a definitive detection of PM. 
We attribute this null result to several compounding factors. 
The two datasets were obtained with different telescopes, configurations, and phase calibrators, any of which can introduce systematic astrometric shifts at the level of $\sim$0.1~beam widths \citep{Masque2015}; while our coordinate alignment via source C mitigates global offsets, it cannot fully correct for differential, direction-dependent phase errors across the field.
Furthermore, the temporal baseline of $\Delta t \approx 1.92$~yr is, even for the fastest component (OW, $64 \pm 12$~mas~yr$^{-1}$), too short as the predicted displacement amounts to only $\sim$123~mas.
Also, the extended, asymmetric morphologies of the outer lobes mean that apparent peak shifts are more likely due to $uv$-coverage differences than genuine motion.
We therefore conclude that the inherent differences between the two datasets preclude a statistically meaningful PM measurement, and defer this analysis to future observations with a longer temporal coverage and a matched observational setup. 

\section{Discussion}\label{discussion}

\subsection{Episodic Accretion and Ejection}\label{episodic}

As mentioned above, outflows and jets are intrinsically connected to the accretion of matter onto the protostar, and therefore the
accretion history is coded in the structure and appearance of the outflowing matter. While early models assumed
steady accretion powering the flows, there is now a large body of evidence for episodic accretion and subsequent ejection as a major factor in the formation of high-mass stars \citep[e.g.,][]{Burns2023, Cesaroni2018}.

Therefore, considering the knotty morphology of the I16562 jet as an indication of accretion/ejection events, we can estimate the time between
events from the dynamical time scale corresponding to each radio component. Since we do not detect significant proper motion in our 2017 VLA observations (Section~\ref{proper_motion}), we adopt the plane-of-sky velocities $v_s$ measured by \citet{e_guzman_2016} at 9~GHz. 
Using these velocities and the projected angular separations from source C-2, and assuming that the shocks traveled with constant velocity, we compute dynamical timescales for each component
via $\tau_\mathrm{dyn} = d / v_\mathrm{s}$, where $d$ is the projected distance from the central source. 
The resulting timescales are $\tau_\mathrm{dyn} = 615 \pm 150$, $92 \pm 27$, $196 \pm 100$, and $339 \pm 65$~yr for the OE, IE, IW, and OW lobes, respectively, indicating that at least 3 accretion events occurred in the last $\sim 600\,$yr. However, it is clear that this number is a lower limit due to the use of the low angular resolution data, as well as the evidence discussed below for ejection events which have no radio counterpart (see Section \ref{sec:precession}).

Our new data at a much higher angular resolution at C-band, which is $\sim$ $0\farcs3$ along the jet direction, allows to characterize the ejection history more finely. In Table~\ref{tab:bigtable} we report 19 compact components which could be interpreted as individual shocks due to ejection events.
In this case a rough estimate for the average time period between ejection events would be 32\,yr. This number would be an upper limit for the above mentioned reasons, and additionally, we note when identifying compact components in the radio continuum emission there
appear to exist several more sources just below our adopted sensitivity limit.

Of particular interest in this discussion is the interior structure of the OE components. As can be seen in Figure~\ref{fig:regions_grid_outer}, upper left panel,
there are 6 compact components which are located approximately equidistant along the east-west direction. Assuming that these components
move with the measured sky velocity of \citet{e_guzman_2016}, the average time between ejections would be approximately $\sim19\,$yr.
Finally, it is also worth mentioning that an equidistant shock structure as seen for the OE subcomponents has been interpreted as 
indication of instabilities in the jet speed due additional sinusoidal velocity components \citep[e.g.,][]{Nony2020, Lora2024}. 

\subsection{Nature of the Flow}
\label{sec:precession}

\begin{figure*}
    \centering
    \includegraphics[width=1\linewidth]{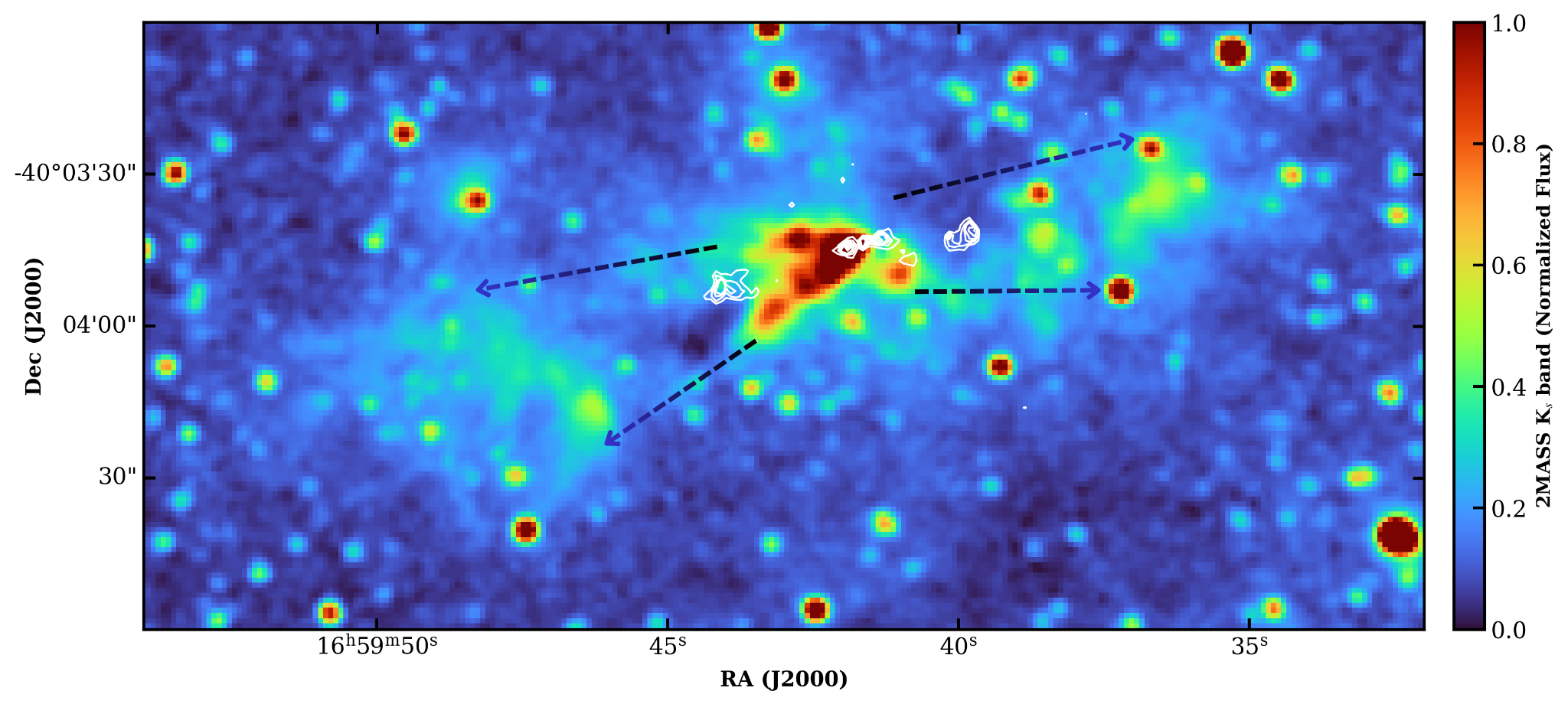}
    \caption{2MASS $\text{K}_s$-band $2.19 \ \mu$m image of the region surrounding I16562, with the $6.1$~GHz contours overlaid in white. The wide angle outflow walls are traced with black arrows. See footnote~\ref{fn:2mass} for data source.}
    \label{fig:2mass}
\end{figure*}

\begin{figure*}
    \centering
    \includegraphics[width=1\linewidth]{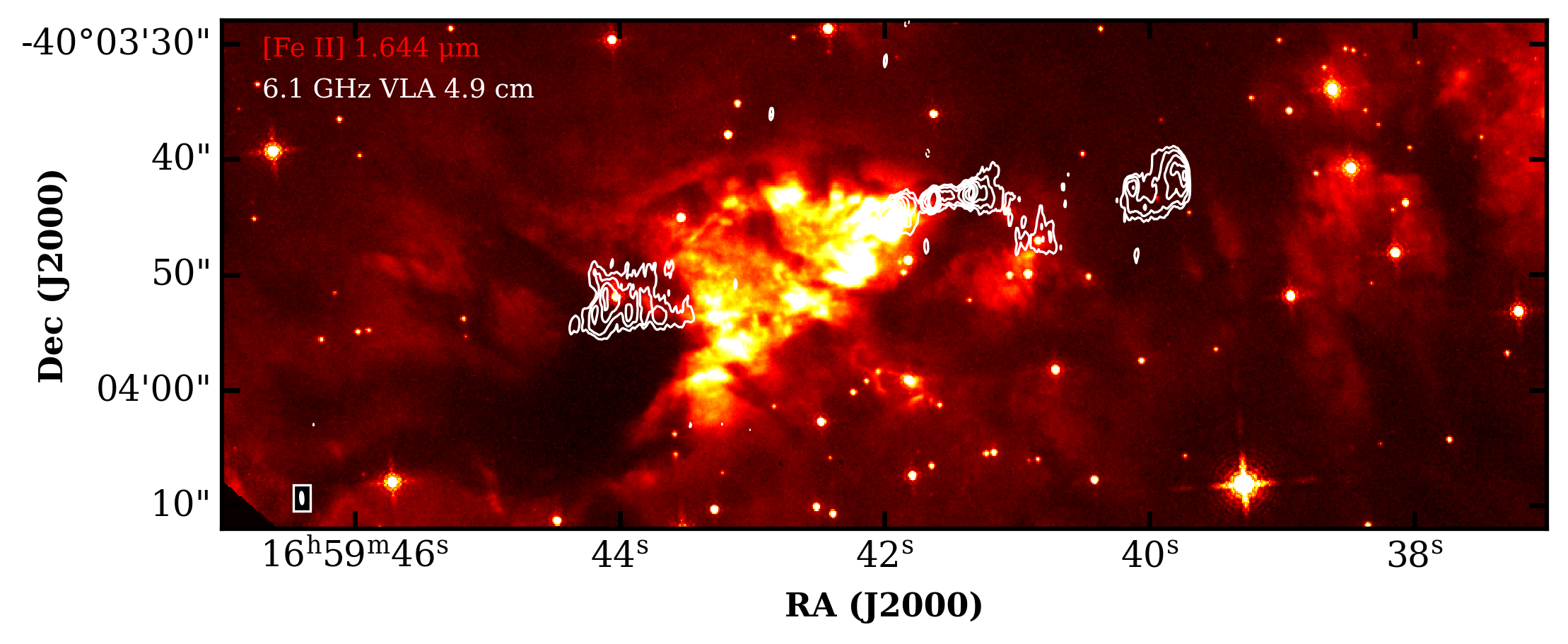}
    \caption{The HST \ce{[FeII]} $1.6 \ \mu$m filter image of I16562, with our 6.1~GHz VLA contours overlaid in white. The flux is normalized over the image. There is a bright, extended, wide outflow extending along the eastern arm of the jet, and both the outer lobes are obscured by dark clouds. See footnote~\ref{fn:hst} for data source.}
    \label{fig:hst_FeII}
\end{figure*}

The two arms of the jet in I16562 are not exactly anti-symmetric: Both the eastern and western outer lobes are offset to the south, forming an angle of about $175^\circ$, i.e. the jet deviates slightly from strict bipolarity.
This systematic southward offset for both outer lobes motivates a discussion of the nature of the large-scale lobe geometry
of the I16562 flow.

Most outflows or jets deviate from exact symmetry, and in some cases the lobes show a distinct S-shape, indicative of precession.
Jet precession is likely caused by the wobble of the inner accretion disk around the central protostar, which produces anti-symmetric angular shifts along both arms \citep[e.g.,][]{Terquem1999, Papaloizou1995}. 
The fact that both outer lobes are displaced in the same direction — southward — rather than in opposite azimuthal directions appears therefore inconsistent with precession as the primary driver of the observed geometry. However, in an attempt to explain the equidistant knots
in the OE component we have considered a precession model where the jet rotates toward the south. Since there is no evidence for a northward
motion of the OW lobe, we would need to additionally require that the jet ejections have occurred in only one direction at a time. 
While monopolar jet emission is a distinct possibility \citep[e.g.,][]{Dutta2026, FendtSheikhnezami2013}, this model appears incompatible with the data of the dynamical timescales, and the proper motion of the jet arms.

We instead favor an interpretation in which the collimated thermal jet is embedded within a wide-angle outflow component, and that the
radio continuum lobe geometry is determined by the surrounding matter.
\citet{RodriguezKamenetzky2022} propose that in the Serpens Triple Radio Source both a collimated jet and a wide-angle wind are simultaneously active, likely launched by a combination of X- and disk-wind mechanisms, and we suggest an analogous two-component structure may be present in I16562. 

Three independent lines of evidence support the presence of such a wide-angle outflow component.
First, the 2MASS $K_s$-band\footnote{\label{fn:2mass}The $K_s$-band Atlas Image was obtained from the NASA/IPAC Infrared Science Archive (IRSA) at \url{https://irsa.ipac.caltech.edu/applications/2MASS/IM/interactive.html}.} image of I16562 (Figure~\ref{fig:2mass}) provides additional information on the large-scale extent of the outflow and its age. 
The $K_s$-band emission at $2.19\,\mu$m traces a combination of shocked $\mathrm{H}_2$ emission and scattered continuum light escaping from the outflow cavity \citep[e.g.,][]{Bally2016}. 
Figure~\ref{fig:2mass} shows bright V-shaped emission toward the east/southeast with its apex located at component C. The opening angle of this structure is $36^\circ$, and it is only seen on one side, presumably due to extinction. The ionized jet is located close
to the geometrical center axis of the V-shaped $K_s$-band emission.  At larger scales, the
2MASS image reveals extended $K_s$-band emission oriented perpendicular to the jet direction, consistent with an extrapolation of the 
central V-shaped structure. The distance between the outer $K_s$-band emission is $\sim 130''$ — corresponding to a physical scale of $\sim 1.1$\,pc at the 1.8\,kpc — far beyond the 0.35\,pc extent of the ionized jet detected in the observed radio wavelengths. 
This extended $K_s$ emission was also noted by \citet{Guzman2011}, who remarked that it appears to trace outflow activity older than the dynamical timescale of the radio-detected outer lobes. 
If the termination of the $K_s$-band flow is used as an outer boundary for the outflow and combined with the proper motion velocities of the outer lobes \citep{e_guzman_2016}, the implied dynamical age of the large-scale outflow is $\sim 1{,}400$\,yr, roughly twice the $\sim 615$\,yr dynamical age of OE derived from the radio-detected emission alone (Section~\ref{episodic}). 
In Figure~\ref{fig:2mass} we can also see that the outer radio lobes are located at the edge of dark clouds, hence they likely arise when the fast jet encounters dense material and strong shocks are formed.

Second, in Figure~\ref{fig:hst_FeII} we show an archival  Hubble Space Telescope (HST)  $1.64 \ \mu$m \ce{[FeII]} narrow-band\footnote{\label{fn:hst} HST GO Program 14465 (PI: J.~Tan, ``Peering to the Heart of Massive Star Birth,'' Cycle~23); \url{https://archive.stsci.edu/proposal_search.php?mission=hst&id=14465}.} image of I16562.
This forbidden line is an indicator of shocked gas in outflows \citep[e.g.][]{Mohan2023}, and particularly traces low ionization gas
in fast J-type shocks (e.g Hollenbach \& McKee 1989).
An inspection of Figure~8 reveals a wide-angle morphology in the near-infrared that is not apparent in the radio. The IR emission traces a one-sided cone-like outflow cavity, similar to the V-shaped emission discussed above. The opening angle of the $1.64 \ \mu$m \ce{[FeII]}
is $38^\circ$, nearly identical to the structure observed in the $K_s$-band. 
This large-scale structure in the near-infrared tracing shocked and excited gas along the cavity walls is best explained by the action of a wide-angle wind sweeping up and deflecting ambient material. Furthermore, since the cooling time of the $1.64 \ \mu$m \ce{[FeII]}
line is rather short \citep[e.g.][]{AllenBurton1993}, outflow activity must still be ongoing.

The HST image shown in Figure~\ref{fig:hst_FeII} also reveals that the outer radio lobes OE and OW are spatially coincident with regions of enhanced extinction visible as dark patches in the \ce{[FeII]} image, suggesting the presence of dense molecular cloudlets along the jet path. 
The interaction of the collimated jet with these clumps of matter would produce strong shocks at the jet--cloud interface, contributing to the shock ionization observed in the outer lobes, and causing the radio emission.
Dense cloudlets along the propagation path would further modulate the shock structure, explaining the irregular brightness distribution and morphological complexity we observe across the OE and OW sub-components.

Third, is the presence of the non-thermal source MW and its morphological and temporal properties. 
Source MW is clearly off-axis southward from the western jet arm, and shows evidence for flux density variability across epochs (see Section~\ref{variability}). 
Since there is no corresponding counterpart to source MW in the eastern jet arm, this source is further evidence of the inhomogeneity of the ambient medium in the jet.
The continued brightening of MW is consistent with the emergence of a compact, shock-ionized knot excited by a relatively fast stellar wind rather
than caused by the collimated jet.

\subsection{The Non-Thermal Source MW}
\label{subsec:mw}

Among all radio sources detected in our VLA observations of I16562, source~MW is unique in displaying a spectral index of $\alpha = -0.7$ (where $S_\nu \propto \nu^\alpha$), steeper than the $\alpha \approx -0.1$ expected for optically thin thermal free-free emission. 
The corresponding power-law index of the electron energy distribution, $p = 1 - 2\alpha = 2.4$, is consistent with relativistic particle acceleration in a non-relativistic shock via the diffusive shock acceleration (DSA) mechanism \citep[]{Bell1978}. 
We therefore identify the emission from MW as synchrotron radiation produced by a population of shock-accelerated relativistic electrons.

To characterize the energetics of the synchrotron-emitting plasma, we apply the minimum energy (equipartition) formalism following \citet{Pacholczyk1970}, as adapted by \citet{CarrascoGonzalez2010} and \citet{Rodriguez2008}. 
The magnetic field strength that minimizes the total energy, $B_{\mathbf{me}}$, and the corresponding total energy, $E_{\mathbf{me}}$, are given by
\begin{equation}
    B_{\mathbf{me}} = 4.5^{2/7}(1+k)^{2/7} c_{12}^{2/7}\phi^{-2/7} R^{-6/7} L_R^{2/7},
    \label{eq:Bmin}
\end{equation}
\begin{equation}
    E_{\mathbf{me}} = c_{13}(1+k)^{4/7} \phi^{3/7}R^{9/7} L_R^{4/7},
    \label{eq:Emin}
\end{equation}
where $k = 40$ is the ratio of energy in heavy particles to that in electrons \citep{BeckKrause2005}, $\phi = 0.5$ is the volume filling factor, $R$ is the physical radius of the source in cm, $L_R$ is the specific radio luminosity in erg~s$^{-1}$~Hz$^{-1}$, and $c_{12} = 1.558 \times 10^7$ and $c_{13} = 1.858 \times 10^4$ are tabulated constants from \citet{Pacholczyk1970}.
Here $E_{\mathbf{me}}$ is the sum of the energy stored in the magnetic field and in the relativistic particles (electrons and heavy particles) within the emitting volume \citep*[e.g.,][Fig.~9.6]{Snell2019Book}.
The angular radius of MW is \textbf{$\theta_s = 2\farcs15$}, corresponding to a physical radius of $R \approx 5.5 \times 10^{16} \ \mathrm{cm}$ at $d = 1.8\ \mathrm{kpc}$.

Substituting these values, we obtain a magnetic field strength of $B_{\mathbf{me}} \approx 0.4\ \mathrm{mG}$ and a total energy of $E_{\mathbf{me}} \approx 4.4 \times 10^{42}\ \mathrm{erg}$.
The corresponding number density of relativistic electrons $n_{er}$ is estimated following \citet{Garay1996} as
\begin{equation}
\begin{split}
    \frac{n_{er} \cdot E_\mathrm{low}}{10^{-9}\ \mathrm{erg\,cm}^{-3}}
    = 3.58
    \left(\frac{S_\nu}{\mathrm{mJy}}\right)
    \left(\frac{\theta_s}{\mathrm{arcsec}}\right)^{-3}
    \left(\frac{\nu}{10\ \mathrm{GHz}}\right)^{1/2} \\
    \times
    \left(\frac{D}{\mathrm{kpc}}\right)^{-1}
    \left(\frac{B}{\mathrm{mG}}\right)^{-3/2},
\end{split}
    \label{eq:ner}
\end{equation}
where $E_{\mathbf{low}} = m_ec^2$ is the low-energy cutoff of the relativistic electron energy distribution, yielding $n_{er} \approx 1 \times 10^{-4}\ \mathrm{cm}^{-3}$. 
This value seems low compared to similar calculations for the synchrotron sources in ionized jet IRAS 16547$-$4247 in \citet{Rodriguez2008} due to the cubed inverse relation with angular size of the source. Therefore we note that this value is likely a lower limit because the synchrotron emitting source size is not fully resolved, and there may be multiple smaller synchrotron sources within MW.

We can also compare the energy of the synchrotron source with the total energy of the jet.
Using the total momentum injected by the jet's bow shocks over its dynamical time, $P_j \approx 9 - 12 \ M_\odot\,\mathrm{km\,s^{-1}}$\textbf{, as derived by \citet{e_guzman_2016}}, and the jet velocity range $V_j \approx 460-1050 \, \mathrm{km\,s^{-1}}$ bracketed by the proper motion and escape velocity constraints of \citet{e_guzman_2016}, we estimate a total jet kinetic energy of $E_\mathrm{kin,jet} \boldsymbol{\gtrsim} 4-12 \times 10^{46}$ erg, a lower limit since $P_j$ is the fraction of the jet momentum transferred to the ambient medium.
The minimum energy stored in the synchrotron-emitting plasma, $E_{\mathbf{me}} \approx 4.4 \times 10^{42}\ \mathrm{erg}$, thus represents a small fraction of this budget, suggesting that while MW marks an energetically significant site of particle acceleration, the bulk of the jet's mechanical power is carried in its thermal, non-relativistic flow.

In section \ref{variability} above we have shown that the MW source has brightened by nearly a factor of 2 in a period of about $2\,$yr. 
This short-timescale variability likely reflects an intrinsic change in the source, most plausibly the emergence of a newly ejected knot or the re-energization of an existing shock. 
In order to better understand the cause of the variability further observations will be required. 
We note that the synchrotron cooling time in regions with such weak B-fields is of order $10^4\,$yr \citep*[e.g.][Eq.~9.11]{Snell2019Book}. Specifically, $\tau_\mathrm{synch} \approx 0.86\, \mathrm{yr}\, (\nu/\mathrm{GHz})^{-1/2} (B/\mathrm{G})^{-3/2}$. For $B=B_{\mathbf{me}}=0.4\,$mG, this gives $\tau_\mathrm{synch} \approx(4\text{--}6) \times 10^{4}\,$yr at 3.1--6.1\,GHz, for emitting electrons with energies $E \approx 1.3\text{--}1.8\,$GeV \citep*[e.g.,][Eq. 3.27]{Snell2019Book}.  
Hence, any weakening of the source based on this effect is not expected.
However, it is likely that MW consists of a mixture of thermal and non-thermal emission, and it is possible that the thermal gas also contributes to the variability. 

\section{Summary} \label{sec:summary}

We have presented new high-resolution, high-sensitivity VLA observations toward the high-mass protostellar jet system IRAS~16562$-$3959.
Our C-band (6.1~GHz) observations achieve a synthesized beam of $1\farcs03 \times 0\farcs26$ with an rms noise of $\sigma = 6.2~\mu$Jy~beam$^{-1}$, and our S-band (3.1~GHz) observations achieve a beam of $2\farcs10 \times 0\farcs48$ with $\sigma = 31~\mu$Jy~beam$^{-1}$. 
Together, these provide one of the most detailed radio continuum pictures of an ionized high-mass protostellar jet yet obtained. Our main results are summarized as follows.

\begin{enumerate}

\item Our VLA images recover the five previously known jet components (OE, IE, C, IW, OW), resolving each large-scale lobe into individual knots and sub-components. 
In total, we detect 21 distinct radio sources associated with the ionized jet. 
We also recover three field sources previously reported by \citet{e_guzman_2016} at higher frequencies and report the discovery of a new compact source, Source~4, at $\alpha_\mathrm{J2000} = 16^{\mathrm{h}}59^{\mathrm{m}}41\fs135$, $\delta_\mathrm{J2000} = -40\degr02\arcmin59\farcs10$, with $S_\nu = 97.0 \pm 4.8~\mu$Jy at 6.1~GHz. 
Based on its compact morphology, association with an X-ray source from \citet{montes_chandra_2020}, and absence of known infrared or maser counterparts, Source~4 is consistent with a T~Tauri star.

\item We report the detection of a new continuum component MW located about  $10^{\prime\prime}$ to the south-west of the center of the flow, i.e., south of the ionized jet axis. MW has a steep negative spectral index of $\alpha \approx -0.7$ between 3.1 and 6.1~GHz, suggesting synchrotron emission driven by shock-accelerated relativistic electrons. 
Applying the minimum-energy (equipartition) formalism, we obtain values for $B_\mathrm{min} \approx 0.4$~mG and $E_\mathrm{min} \approx 4.4 \times 10^{42}$~erg, which allows us to put a lower limit on the corresponding relativistic electron number density of $n_{e,r} \approx 1 \times 10^{-4}$~cm$^{-3}$. 

\item We do not detect any significant flux density variability in the previously known continuum components. 
However, the newly identified non-thermal source MW has brightened substantially relative to its marginal detections in the earlier ATCA data. 
Furthermore we were not able to  detect statistically significant proper motion of any jet component, which we attribute to the short temporal baseline of
$\Delta t \approx 1.92$~yr between our VLA epoch and the 2015 ATCA observations, combined with different $uv$-coverage, and different configurations between the two datasets.

\item The dynamical timescales of the jet lobes, estimated from the lobe separations and the proper motions reported by \citet{e_guzman_2016}, span $\sim$90--615~yr from inner to outer lobes. If we interprete the 19 compact components in the jet lobes as individual shocks due to ejection events, we
estimate that the average time period between ejection events would be 32 yr.

\item A 2MASS $K_s$-band image of I16562 shows extended emission over $\sim 130''$ ($\sim$1.1~pc at 1.8~kpc), well beyond the $\sim$0.35~pc extent of the radio-detected ionized jet. 
If the termination of this $K_s$-band emission is adopted as the outer boundary of the outflow and combined with the proper motion velocities of \citet{e_guzman_2016}, the implied dynamical age of the large-scale outflow is $\sim$1,400~yr. 
The $K_s$-band emission has a cone-like geometry with an opening angle of $36^\circ$, with its apex located near component C.
This geometry is also observed in a HST [\ce{FeII}]\,1.64~$\mu$m image, which reveals the presence of low-ionization gas in the gap between
components IE and OE. The outer radio components OE and OW are observed at the edge of dark clouds in both images, suggesting that the ionization occurs
when the fast jet interacts with dense material along its path. Together these finding strongly suggest that 
the collimated ionized jet in IRAS~16562$-$3959 is embedded within a larger-scale wide-angle wind.

\end{enumerate}

Taken together, our multi-frequency VLA observations, along with archival NIR data, provide a detailed picture of the morphological, kinematic, and temporal properties of the ionized jet I16562 driven by the HMYSO G345.4938+01.4677 at its core. 
The data are consistent with a shock-ionized protostellar jet embedded in a wider-angle wind and driven by a high-mass protostar undergoing variable, episodic accretion. 
This work also demonstrates the scientific value of targeting low-declination southern star-forming regions with high-resolution, high-sensitivity northern-hemisphere facilities such as the VLA, despite the observational challenges imposed by low target elevation. 
A follow-up campaign approximately a decade after our 2017 VLA observations would enable a statistically significant proper motion measurement, tighter constraints on the variability of MW, and direct monitoring of structural changes within the resolved knot substructure.

\section*{Acknowledgements}

Support for this work was partially provided by the NSF Fund for the NRAO Graduate Summer Student Research Program.

The National Radio Astronomy Observatory is a facility of the National Science Foundation operated under cooperative agreement by Associated Universities, Inc. 

A.S. thanks the LSST-DA Data Science Fellowship Program, which is funded by LSST-DA, the Brinson Foundation, the WoodNext Foundation, and the Research Corporation for Science Advancement Foundation; his participation in these programs has benefited this work.

\bibliographystyle{aasjournal}
\bibliography{bib2}
\end{document}